\documentclass[aps,prd,twocolumn,showpacs,groupedaddress]{revtex4}
\usepackage{graphicx}
\usepackage{subfigure}
\usepackage{booktabs}
\usepackage[colorlinks,linkcolor=black,urlcolor=blue,citecolor=blue,bookmarks,bookmarksnumbered]{hyperref}

\begin{document}

\title{Quark-lepton complementarity and self-complementarity in different schemes}

\newcommand*{\PKU}{School of Physics and State Key Laboratory of Nuclear Physics and Technology, Peking University, Beijing 100871,
China}\affiliation{\PKU}
\newcommand*{\CHEP}{Center for High Energy Physics, Peking University, Beijing 100871, China}\affiliation{\CHEP}

\author{Yijia Zhang}\affiliation{\PKU}
\author{Xinyi Zhang}\affiliation{\PKU}
\author{Bo-Qiang Ma}\email{mabq@pku.edu.cn}\affiliation{\PKU}\affiliation{\CHEP}

\begin{abstract}
With the progress of increasingly precise measurements on the
neutrino mixing angles, phenomenological relations such as
quark-lepton complementarity~(QLC) among mixing angles of quarks and
leptons and self-complementarity~(SC) among lepton mixing angles
have been observed. Using the latest global fit results of the quark
and lepton mixing angles in the standard Chau-Keung scheme, we
calculate the mixing angles and {\it CP}-violating phases in the other
eight different schemes. We check the dependence of these mixing
angles on the {\it CP}-violating phases in different phase schemes. The
dependence of QLC and SC relations on the {\it CP} phase in the other
eight schemes is recognized and then analyzed, suggesting that
measurements on {\it CP}-violating phases of the lepton sector are crucial
to the explicit forms of QLC and SC in different schemes.
\end{abstract}


\maketitle

\section{\label{sec1}INTRODUCTION}
After decades of neutrino oscillation experiments, it is generally
taken for granted that neutrinos are massive particles that can vary
among all the three flavors through the oscillation process described by
neutrino mixing. One of the most important issues concerning
neutrino mixing is the determination of the neutrino mixing matrix,
i.e., the Pontecorvo-Maki-Nakagawa-Sakata~(PMNS) matrix~\cite{PMNS},
which is the lepton sector counterpart of the quark sector mixing
matrix, the Cabibbo-Kobayashi-Maskawa~(CKM) matrix~\cite{CKM}. The
PMNS matrix is defined as the correlation matrix linking neutrino
flavor eigenstates $|\nu_{\rm flavor}\rangle$ and mass eigenstates
$|\nu_{\rm mass}\rangle$,
\begin{equation}
      |\nu_{\rm mass}\rangle=U_{\rm PMNS}|\nu_{\rm flavor}\rangle.
\end{equation}
This mixing matrix is conventionally represented in the standard
Chau-Keung (CK) scheme~\cite{CK} as
\begin{widetext}
    \begin{equation}\label{CKe}
      U_{\rm PMNS}=\left(\begin{array}{ccc}
        c_{12}c_{13} & s_{12}s_{13} & s_{13}e^{-i\delta}\\
        -s_{12}c_{23}-c_{12}s_{23}s_{13}e^{i\delta} & c_{12}c_{23}-s_{12}s_{23}s_{13}e^{i\delta} & s_{23}c_{13}\\
        s_{12}s_{23}-c_{12}s_{23}s_{13}e^{i\delta} & -c_{12}s_{23}-s_{12}s_{23}s_{13}e^{i\delta} & c_{23}c_{13}
      \end{array}\right)
      \left(\begin{array}{ccc}
        1 & & \\
         & e^{i\alpha} & \\
         & & e^{i\beta}
      \end{array}\right),
    \end{equation}
\end{widetext}
where three mixing angles are denoted by $\theta_{12}$,
$\theta_{23}$, and $\theta_{13}$, with their trigonometric functions
$\sin\theta_{12}$, $\cos\theta_{12}$, etc. represented by $s_{12}$,
$c_{12}$, etc. respectively. The {\it CP}-violating phase is denoted by
$\delta$, meanwhile $\alpha$ and $\beta$ represent the other two
phases in the case of Majorana neutrinos. In the case of Dirac
neutrinos, the latter two phases $\alpha$ and $\beta$ can be removed
by redefinition, thus there remain only four independent parameters,
i.e., three mixing angles together with one {\it CP}-violating phase. If
the neutrinos are of Majorana type, the two phases $\alpha$ and
$\beta$ are needed for a full determination of the mixing matrix.
As the Majorana phases do not manifest themselves in the
oscillation, we ignore these two phases $\alpha$ and $\beta$ and
take only the first term on the right-hand side of Eq.~(\ref{CKe}) in
this article. By now, the quark-sector mixing matrix has been measured with good
precision. In the lepton sector, the values of the three mixing
angles have been measured after years of neutrino oscillation
experiments, though with relatively lower precision compared to the
quark case.

The explicit form of the fermion mixing matrix is not unique and an
alternative scheme is the Kobayashi-Maskawa~(KM) scheme~\cite{KM}
    \begin{eqnarray}\label{KMe}
      &&U_{\rm PMNS}\\
      &&=\left(\begin{array}{ccc}
            c_1 & s_1c_3 & -s_1s_3\\
            -s_1c_2 & c_1c_2c_3+s_2s_3e^{-i\delta} & -c_1c_2s_3+s_2c_3e^{-i\delta}\\
            s_1s_2 & -c_1s_2c_3+c_2s_3e^{-i\delta} & c_1s_2s_3+c_2c_3e^{-i\delta}
      \end{array}\right).\nonumber
    \end{eqnarray}
In the KM scheme, as will be mentioned later, the {\it CP}-violating phase
of the quark sector is quite near $90^\circ$, leading to the
hypothesis of ``maximal {\it CP}
violation''~\cite{koide,Koide:2008yu,boomerang,Li:2010ae,qinnan,Ahn:2011it}.
Besides the CK and KM schemes, in Ref.~\cite{nine} all the other possible
schemes of the mixing matrix are considered and presented. There are actually
12 schemes of mixing matrix. Among them, 3 schemes can be transformed into
others through straightforward redefinition of mixing angles, thus leaving 9
different schemes~\cite{nine,Zheng10,Zhang:2012xu,QLCnine}, whose forms are
provided in Sec.~\ref{sec2}.

Quark-Lepton
Complementarity~(QLC)~\cite{smirnov,qlc,raidal,phenomenology} and
Self-Complementarity~(SC)~\cite{Zheng:2011uz,Zhang:2012xu} are
phenomenological relations of quark and lepton mixing angles. They
provide a novel connection to link quarks and leptons. They can be
expressed in a more clear way as (\ref{QLCSC1})-(\ref{QLCSC4}):
\begin{eqnarray}
      & \mbox{QLC1: } & \theta_{12}+\vartheta_{12}=45^\circ, \label{QLCSC1}\\
      & \mbox{QLC2: } & \theta_{23}+\vartheta_{23}=45^\circ, \\
      & \mbox{SC1: } & \vartheta_{12}+\vartheta_{13}=45^\circ, \\
      & \mbox{SC2: } & \vartheta_{12}+\vartheta_{13}=\vartheta_{23}. \label{QLCSC4}
\end{eqnarray}
(From now on we use $\vartheta$ to represent lepton sector mixing
angles to distinguish them from quark sector mixing angles
$\theta$.)

Here we have marked the two QLC relations by QLC1 and QLC2, and the
two slightly different SC relations by SC1 and SC2 respectively.
Originally these phenomenological relations are observed only in the
CK scheme and fit the experimental results within small errors.
However, a question naturally arises, i.e., whether these relations
still hold in schemes other than the CK scheme since we cannot find
any justification for the priority of the CK scheme. There are
already some researches on QLC and SC in the nine
schemes~\cite{Zheng10,Zhang:2012xu,QLCnine}. However, all of these
examinations of QLC and SC are carried out under some fixed phase
choices. Since the {\it CP}-violating phase of the lepton sector is not
determined from current experiment, in this article we examine QLC
and SC with the whole range variation of the lepton {\it CP}-violating
phase. These will be treated in Sec.~\ref{sec3} and Sec.~\ref{sec4}.

We purpose to make a detailed re-analysis of QLC and
SC in all the nine schemes, emphasizing on the influences due to the
variation of the lepton {\it CP}-violating phase. In Sec.~\ref{sec2} we
use the latest experiment results to do calculations on mixing
angles and {\it CP}-violating phases in all the nine schemes. In
Sec.~\ref{sec3} we focus on QLC, examine these complementarity
relations and make some analyses. In Sec.~\ref{sec4} we similarly
examine and analyze SC. In Sec.~\ref{sec5} we discuss some
properties of {\it CP}-violating phases among different schemes with a
suggestion of convention redefinition, and suggest some empirical
relations among quark {\it CP}-violating phases in different schemes.

\section{\label{sec2}QUARK AND LEPTON MIXING ANGLES AND CP-VIOLATING PHASES}
    First we list all the nine schemes mentioned in Sec.~\ref{sec1} in Table~\ref{tabscheme}.
\begingroup
\squeezetable
\begin{table*}
      \caption{Nine different schemes of fermion mixing matrix}\label{tabscheme}
      \begin{ruledtabular}
      \begin{tabular}{cc}
      \toprule
      Scheme & Mixing angles and Jarlskog invariant\\
      \hline
      $\underline{P1:U=R_{23}(\theta_{23})R_{31}(\theta_{13},\phi)R_{12}(\theta_{12})}$ & $J_1=s_{12}s_{23}s_{13}c_{12}c_{23}c^2_{13}\sin\phi$\\
      $\left(\begin{array}{ccc}
            c_{12}c_{13} & s_{12}s_{13} & s_{13}\\
            -c_{12}s_{23}s_{13}-s_{12}c_{23}e^{-i\phi} & -s_{12}s_{23}s_{13}+c_{12}c_{23}e^{-i\phi} & s_{23}c_{13}\\
            -c_{12}s_{23}s_{13}+s_{12}s_{23}e^{-i\phi} & -s_{12}c_{23}s_{13}-c_{12}s_{23}e^{-i\phi} & c_{23}c_{13}
      \end{array}\right)$ &
      $\begin{array}{c}
        \theta_{12}=\arcsin\frac{|U_{12}|}{|U_{13}|}\\ \theta_{23}=\arctan\frac{|U_{23}|}{|U_{33}|}\\ \theta_{13}=\arcsin |U_{13}|
      \end{array}$\\
      $\underline{P2:U=R_{12}(\theta_3)R_{23}(\theta_2,\phi)R^{-1}_{12}(\theta_1)}$ & $J_2=s_1s^2_2s_3c_1c_2c_3\sin\phi$\\
      $\left(\begin{array}{ccc}
            s_1c_2s_3+c_1c_3e^{-i\phi} & c_1c_2s_3-s_1c_3e^{-i\phi} & s_2s_3\\
            s_1c_2c_3-c_1s_3e^{-i\phi} & c_1c_2c_3+s_1s_3e^{-i\phi} & s_2c_3\\
            -s_1s_2 & -c_1s_2 & c_2
      \end{array}\right)$ &
      $\begin{array}{c}
        \theta_1=\arctan\frac{|U_{31}|}{|U_{32}|}\\ \theta_2=\arccos |U_{33}|\\ \theta_3=\arctan\frac{|U_{13}|}{|U_{23}|}
      \end{array}$\\
      $\underline{P3:U=R_{23}(\theta_2)R_{12}(\theta_1,\phi)R^{-1}_{23}(\theta_3)}$ & $J_3=s^2_1s_2s_3c_1c_2c_3\sin\phi$\\
      $\left(\begin{array}{ccc}
            c_1 & s_1c_3 & -s_1s_3\\
            -s_1c_2 & c_1c_2c_3+s_2s_3e^{-i\phi} & -c_1c_2s_3+s_2c_3e^{-i\phi}\\
            s_1s_2 & -c_1s_2c_3+c_2s_3e^{-i\phi} & c_1s_2s_3+c_2c_3e^{-i\phi}
      \end{array}\right)$ &
      $\begin{array}{c}
        \theta_1=\arccos |U_{11}|\\ \theta_2=\arctan\frac{|U_{31}|}{|U_{21}|}\\ \theta_3=\arctan\frac{|U_{13}|}{|U_{12}|}
      \end{array}$\\
      $\underline{P4:U=R_{23}(\theta_2)R_{12}(\theta_1,\phi)R^{-1}_{31}(\theta_3)}$ & $J_4=s_1s_2s_3c^2_1c_2c_3\sin\phi$\\
      $\left(\begin{array}{ccc}
            c_1c_3 & s_1 & -c_1s_3\\
            -s_1c_2c_3+s_2s_3e^{-i\phi} & c_1c_2 & s_1c_2s_3+s_2c_3e^{-i\phi}\\
            s_1s_2c_3+c_2s_3e^{-i\phi} & -c_1s_2 & -s_1s_2s_3+c_2c_3e^{-i\phi}
      \end{array}\right)$ &
      $\begin{array}{c}
        \theta_1=\arcsin |U_{12}|\\ \theta_2=\arctan\frac{|U_{32}|}{|U_{22}|}\\ \theta_3=\arctan\frac{|U_{13}|}{|U_{11}|}
      \end{array}$\\
      $\underline{P5:U=R_{31}(\theta_3)R_{23}(\theta_2,\phi)R^{-1}_{12}(\theta_1)}$ & $J_5=s_1s_2s_3c_1c^2_2c_3\sin\phi$\\
      $\left(\begin{array}{ccc}
            -s_1s_2s_3+c_1c_3e^{-i\phi} & -c_1s_2s_3-s_1c_3e^{-i\phi} & c_2s_3\\
            s_1c_2 & c_1c_2 & s_2\\
            -s_1s_2c_3-c_1s_3e^{-i\phi} & -c_1s_2c_3+s_1s_3e^{-i\phi} & c_2c_3
      \end{array}\right)$ &
      $\begin{array}{c}
        \theta_1=\arctan\frac{|U_{21}|}{|U_{22}|}\\ \theta_2=\arcsin |U_{23}|\\ \theta_3=\arctan\frac{|U_{13}|}{|U_{33}|}
      \end{array}$\\
      $\underline{P6:U=R_{12}(\theta_1)R_{31}(\theta_3,\phi)R^{-1}_{23}(\theta_2)}$ & $J_6=s_1s_2s_3c_1c_2c^2_3\sin\phi$\\
      $\left(\begin{array}{ccc}
            c_1c_3 & c_1s_2s_3+s_1c_2e^{-i\phi} & c_1c_2s_3-s_1s_2e^{-i\phi}\\
            -s_1c_3 & -s_1s_2s_3+c_1c_2e^{-i\phi} & -s_1c_2s_3-c_1s_2e^{-i\phi}\\
            -s_3 & s_2c_3 & c_2c_3
      \end{array}\right)$ &
      $\begin{array}{c}
        \theta_1=\arctan\frac{|U_{21}|}{|U_{11}|}\\ \theta_2=\arctan\frac{|U_{32}|}{|U_{33}|}\\ \theta_3=\arcsin |U_{31}|
      \end{array}$\\
      $\underline{P7:U=R_{31}(\theta_3)R_{12}(\theta_1,\phi)R^{-1}_{31}(\theta_2)}$ & $J_7=s^2_1s_2s_3c_1c_2c_3\sin\phi$\\
      $\left(\begin{array}{ccc}
            c_1c_2c_3+s_2s_3e^{-i\phi} & s_1c_3 & -c_1s_2c_3+c_2s_3e^{-i\phi}\\
            -s_1c_2 & c_1 & s_1s_2\\
            -c_1c_2s_3+s_2c_3e^{-i\phi} & -s_1s_3 & c_1s_2s_3+c_2c_3e^{-i\phi}
      \end{array}\right)$ &
      $\begin{array}{c}
        \theta_1=\arccos |U_{22}|\\ \theta_2=\arctan\frac{|U_{23}|}{|U_{21}|}\\ \theta_3=\arctan\frac{|U_{32}|}{|U_{12}|}
      \end{array}$\\
      $\underline{P8:U=R_{12}(\theta_1)R_{23}(\theta_2,\phi)R^{-1}_{31}(\theta_3)}$ & $J_8=s_1s_2s_3c_1c^2_2c_3\sin\phi$\\
      $\left(\begin{array}{ccc}
            -s_1s_2s_3+c_1c_3e^{-i\phi} & s_1c_2 & s_1s_2c_3+c_1s_3e^{-i\phi}\\
            -c_1s_2s_3-s_1c_3e^{-i\phi} & c_1c_2 & c_1s_2c_3-s_1s_3e^{-i\phi}\\
            -c_2s_3 & -s_2 & c_2c_3
      \end{array}\right)$ &
      $\begin{array}{c}
        \theta_1=\arctan\frac{|U_{12}|}{|U_{22}|}\\ \theta_2=\arccos |U_{32}|\\ \theta_3=\arctan\frac{|U_{31}|}{|U_{33}|}
      \end{array}$\\
      $\underline{P9:U=R_{31}(\theta_3)R_{12}(\theta_1,\phi)R^{-1}_{23}(\theta_2)}$ & $J_9=s_1s_2s_3c^2_1c_2c_3\sin\phi$\\
      $\left(\begin{array}{ccc}
            c_1c_3 & s_1c_2c_3-s_2s_3e^{-i\phi} & s_1s_2c_3+c_2s_3e^{-i\phi}\\
            -s_1 & c_1c_2 & c_1s_2\\
            -c_1s_3 & -s_1c_2s_3-s_2c_3e^{-i\phi} & -s_1s_2s_3+c_2c_3e^{-i\phi}
      \end{array}\right)$ &
      $\begin{array}{c}
        \theta_1=\arcsin |U_{21}|\\ \theta_2=\arctan\frac{|U_{23}|}{|U_{22}|}\\ \theta_3=\arctan\frac{|U_{31}|}{|U_{11}|}
      \end{array}$\\
      \bottomrule
      \end{tabular}
      \end{ruledtabular}
\end{table*}
\endgroup
To avoid ambiguities, the explicit forms of the rotation matrices are provided:
\begin{equation}
      R_{12}(\theta_3)=\left(\begin{array}{ccc}
        c_3 & s_3 & 0\\
        -s_3 & c_3 & 0\\
        0 & 0 & 1
      \end{array}\right),
\end{equation}
\begin{equation}
      R_{31}(\theta_3)=\left(\begin{array}{ccc}
        c_3 & 0 & s_3\\
        0 & 1 & 0\\
        -s_3 & 0 & c_3
      \end{array}\right),
\end{equation}
\begin{equation}
      R_{12}(\theta_3,\phi)=\left(\begin{array}{ccc}
        c_3 & s_3 & 0\\
        -s_3 & c_3 & 0\\
        0 & 0 & e^{-i\phi}
      \end{array}\right),
\end{equation}
and the others are similarly defined. From (\ref{CKe}), (\ref{KMe}) and Table~\ref{tabscheme}, P3 is the same as the KM scheme, and P1 is equivalent to the CK scheme. The {\it CP}-violating phases in Table~\ref{tabscheme} are denoted by $\phi$, rather than $\delta$, to remind readers of the slight difference between the CK scheme in (\ref{CKe}) and the P1 scheme in Table~\ref{tabscheme}. Actually, when we identify $\delta$ as $\phi$, i.e., $\delta=\phi$, we get an equation between the PMNS matrices in the CK scheme and the one in the P1 scheme:
\begin{equation}
      \label{CKP1}
      U_{\rm CK}=\left(\begin{array}{ccc}
        1 &&\\
        & e^{i\phi} &\\
        && e^{i\phi}
      \end{array}\right)
      U_{\rm P1}
      \left(\begin{array}{ccc}
        1 &&\\
        & 1 &\\
        && e^{-i\phi}
      \end{array}\right).
\end{equation}
The phase factors in the two matrices in (\ref{CKP1}) can be eliminated by
unphysical phase redefinition of lepton fields in the case of Dirac neutrinos.
In the case of Majorana neutrinos, the phase factor in the matrix to the right
of $U_{\rm P1}$ can be absorbed by redefinition of Majorana phases, while the
phase factors in the matrix to the left of $U_{\rm P1}$ are still eliminated.

\subsection{Quark sector}
We begin our quark-sector calculations with the experimental data of Wolfenstein parameters~\cite{Wolfenstein:1983yz} listed in (\ref{equ3}) from Particle Data Group~\cite{pdg2012}, together with their relations with the four parameters, i.e., three mixing angles and one {\it CP}-violating phase.
\begin{eqnarray}
      &&\sin\theta_{12}=\lambda, \nonumber\\
      &&\sin\theta_{23}=A\lambda^2, \nonumber\\
      &&\sin\theta_{13}e^{i\delta}=\frac{A\lambda^3(\bar{\rho}+i\bar{\eta})\sqrt{1-A^2\lambda^4}} {\sqrt{1-\lambda^2}[1-A^2\lambda^4(\bar{\rho}+i\bar{\eta})]}, \nonumber\\
      &&\lambda=0.22535\pm0.00065, \label{equ3}\\
      &&A=0.811^{+0.022}_{-0.012}, \nonumber\\
      &&\bar{\rho}=0.131^{+0.026}_{-0.013}, \nonumber\\
      &&\bar{\eta}=0.345^{+0.013}_{-0.014}.\nonumber
\end{eqnarray}
From (\ref{equ3}), we easily get mixing angles and {\it CP}-violating phase in P1:
\begin{eqnarray}
      & P1: & \theta_{12}=\left(13.023^{+0.038}_{-0.038}\right)^\circ, \\
         &&\theta_{23}=\left(2.360^{+0.065}_{-0.038}\right)^\circ, \\
         &&\theta_{13}=\left(0.201^{+0.010}_{-0.008}\right)^\circ, \\
         &&\phi_1=\left(69.10^{+2.02}_{-3.85}\right)^\circ.
\end{eqnarray}
Next, the CKM matrix is calculated from the four parameters
\begin{widetext}
\begin{equation}\label{CKM}
      V_{\rm CKM}=\left(\begin{array}{ccc}
        0.97427\pm0.00015 & 0.22535\pm0.00065 & 0.00352^{+0.00018}_{-0.00015}\\
        0.2252\pm0.0006 & 0.97344^{+0.00015}_{-0.00016} & 0.0412^{+0.0011}_{-0.0007}\\
        0.00867^{+0.00027}_{-0.00027} & 0.0404^{+0.0011}_{-0.0006} & 0.999145^{+0.000027}_{-0.000047}
      \end{array}\right).
\end{equation}
\end{widetext}
When referred to matrix elements of the CKM matrix or the PMNS
matrix, we always mean the absolute value of each matrix element in
this article.

The Jarlskog invariant~\cite{jarlskog} is derived:
\begin{eqnarray}
    J&=&\sin\theta_{12}\sin\theta_{23}\sin\theta_{13}\cos\theta_{12}\cos\theta_{23}\cos^2\theta_{13}\sin\phi\nonumber\\
       &=&(2.97^{+0.17}_{-0.18})\times10^{-5}.
\end{eqnarray}
\subsection{Lepton sector}
The lepton sector is dealt with similarly, with the following normal hierarchy~(NH)
global fit data of mixing angles with $1\sigma$ errors in the P1
scheme~\cite{lepglobalfit}:
\begin{eqnarray}
      &&\sin^2\vartheta_{12}=0.307^{+0.018}_{-0.016}, \nonumber\\
      &&\sin^2\vartheta_{23}=0.386^{+0.024}_{-0.021}, \label{normal}\\
      &&\sin^2\vartheta_{13}=0.0241^{+0.0025}_{-0.0025},\nonumber
\end{eqnarray}
which are equivalently
\begin{eqnarray}
      &&\vartheta_{12}=\left(33.65^{+1.11}_{-1.00}\right)^\circ, \nonumber\\
      &&\vartheta_{23}=\left(38.41^{+1.40}_{-1.24}\right)^\circ, \\
      &&\vartheta_{13}=\left(8.93^{+0.46}_{-0.48}\right)^\circ.\nonumber
\end{eqnarray}
The inverse hierarchy global fit data of the P1 scheme mixing angles with $1\sigma$ errors~\cite{lepglobalfit} are
\begin{eqnarray}
      &&\sin^2\vartheta_{12}=0.307^{+0.018}_{-0.018}, \nonumber\\
      &&\sin^2\vartheta_{23}=0.392^{+0.039}_{-0.022}, \label{inverse}\\
      &&\sin^2\vartheta_{13}=0.0244^{+0.0023}_{-0.0025}.\nonumber
\end{eqnarray}
In this article, we only deal with the case of NH,
because the global fit values for inverse hierarchy in
(\ref{inverse}) are quite close to the values for NH
in (\ref{normal}) and thus our choice does not affect the analysis
and conclusions of this article.

The difference from the quark sector is that at present there are no
experimental results on the lepton {\it CP}-violating phase, but the four
parameters are in a combined transformation when changing schemes.
Therefore, to examine QLC and SC relations in the other eight
schemes, it is necessary to choose a value of the lepton {\it CP}-violating
phase. In this article we will not calculate the PMNS matrix under
certain fixed value of the lepton {\it CP}-violating phase, such as the one with
$\phi_3=90^\circ$~\cite{Zhang:2012ys}. Instead, we will carry out
the calculations with the {\it CP}-violating phase in the P3 scheme $\phi_3$
varying almost continuously from $0^\circ$ to $180^\circ$. (From now
on we use $\phi_i$ with a subscript $i$ to denote $\phi$ in the P$i$
scheme.) The results will be provided in tables with
$\phi_3=0^\circ$, $45^\circ$, $90^\circ$, $135^\circ$, $180^\circ$,
respectively, and in figures by smooth curves. To explicitly show
our methods of calculation, we then carry out the calculations in
detail in the case of the phase $\phi_3=90^\circ$.

First, we calculate the absolute values of five elements of the PMNS
matrix that are independent of the lepton {\it CP}-violating phase from
(\ref{normal}):
\begin{eqnarray}
      &&|U_{11}|=0.822^{+0.010}_{-0.011},\\
      &&|U_{12}|=0.547^{+0.016}_{-0.015},\\
      &&|U_{13}|=0.155\pm0.008,\\
      &&|U_{23}|=0.614^{+0.019}_{-0.017},\\
      &&|U_{33}|=0.774^{+0.013}_{-0.015}.
\end{eqnarray}
Then, the condition $\phi_3=90^\circ$ is used to determine mixing
angles in the P3 scheme:
\begin{eqnarray}
      \cos\vartheta_1&=&|U_{11}|\nonumber\\
      \to\vartheta_1&=&\left(34.68^{+1.07}_{-0.97}\right)^\circ, \\
      \tan\vartheta_3&=&\frac{|U_{13}|}{|U_{12}|}\nonumber\\
      \to\vartheta_3&=&\left(15.83^{+0.88}_{-0.93}\right)^\circ, \\
      |U_{23}|^2&=&\frac{c^2_1s^2_3+c^2_3}{2}+\frac{c^2_1s^2_3-c^2_3}{2}\times\nonumber\\
      &&\cos(2\vartheta_2+\arctan\frac{2c_1s_3c_3\cos\phi_3} {c^2_1s^2_3-c^2_3})\nonumber\\
      \to\vartheta_2&=&\left(37.64^{+1.58}_{-1.40}\right)^\circ.
\end{eqnarray}
Next, we get all the elements of the PMNS matrix:
\begin{equation}
      U_{\rm PMNS}=\left(\begin{array}{ccc}
        0.822^{+0.010}_{-0.011} & 0.547^{+0.016}_{-0.015} & 0.155\pm0.008\\
        0.451^{+0.014}_{-0.014} & 0.648^{+0.012}_{-0.014} & 0.614^{+0.019}_{-0.017}\\
        0.347^{+0.016}_{-0.014} & 0.529^{+0.015}_{-0.014} & 0.774^{+0.013}_{-0.015}
      \end{array}\right),
\end{equation}
together with the Jarlskog invariant
\begin{equation}
      J=0.0338^{+0.0017}_{-0.0018}.
\end{equation}

\begin{figure*}
        \centering
        \subfigure[~P1]{
          \includegraphics[width=5.8cm]{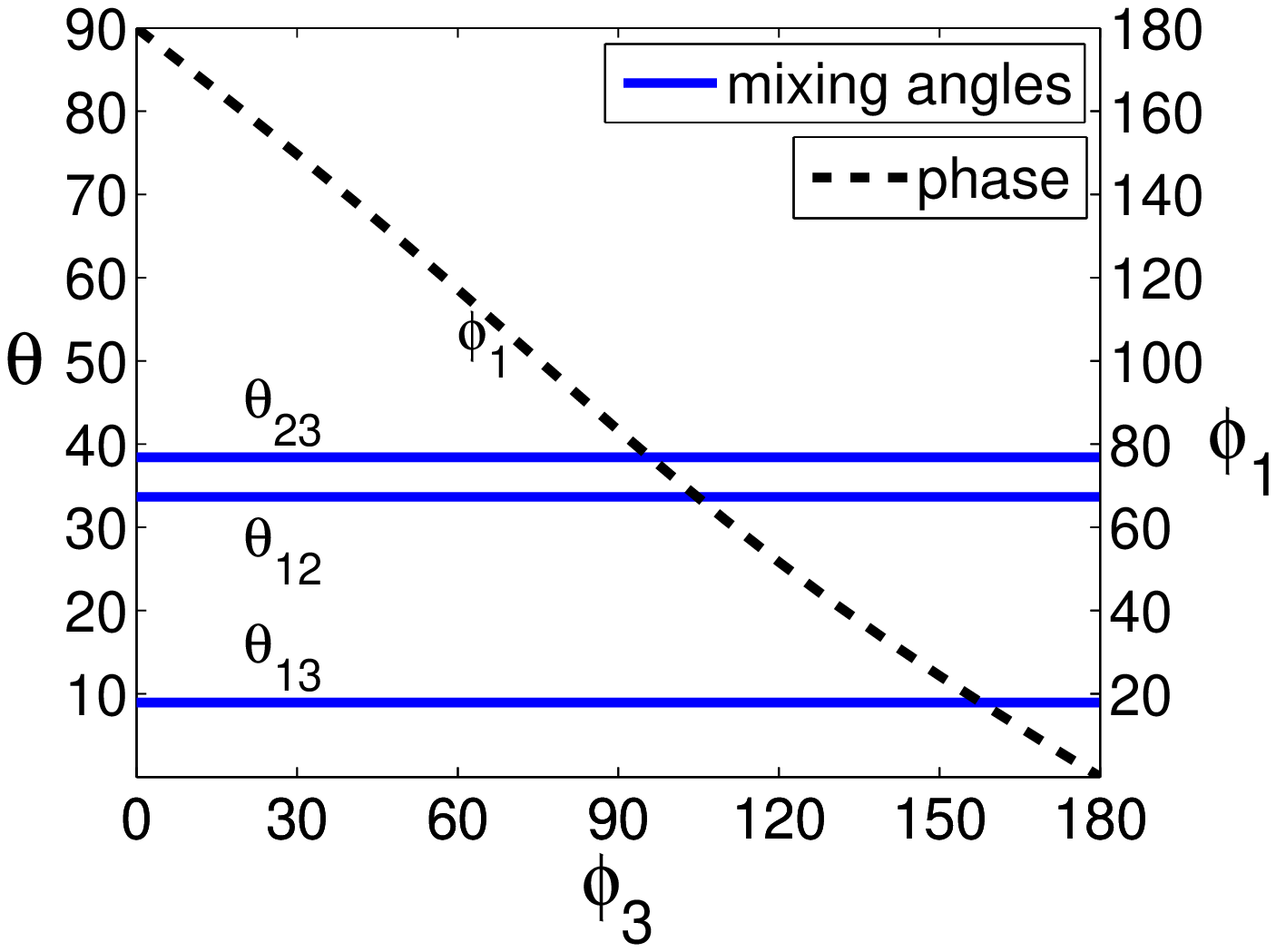}}
        \subfigure[~P2]{
          \includegraphics[width=5.8cm]{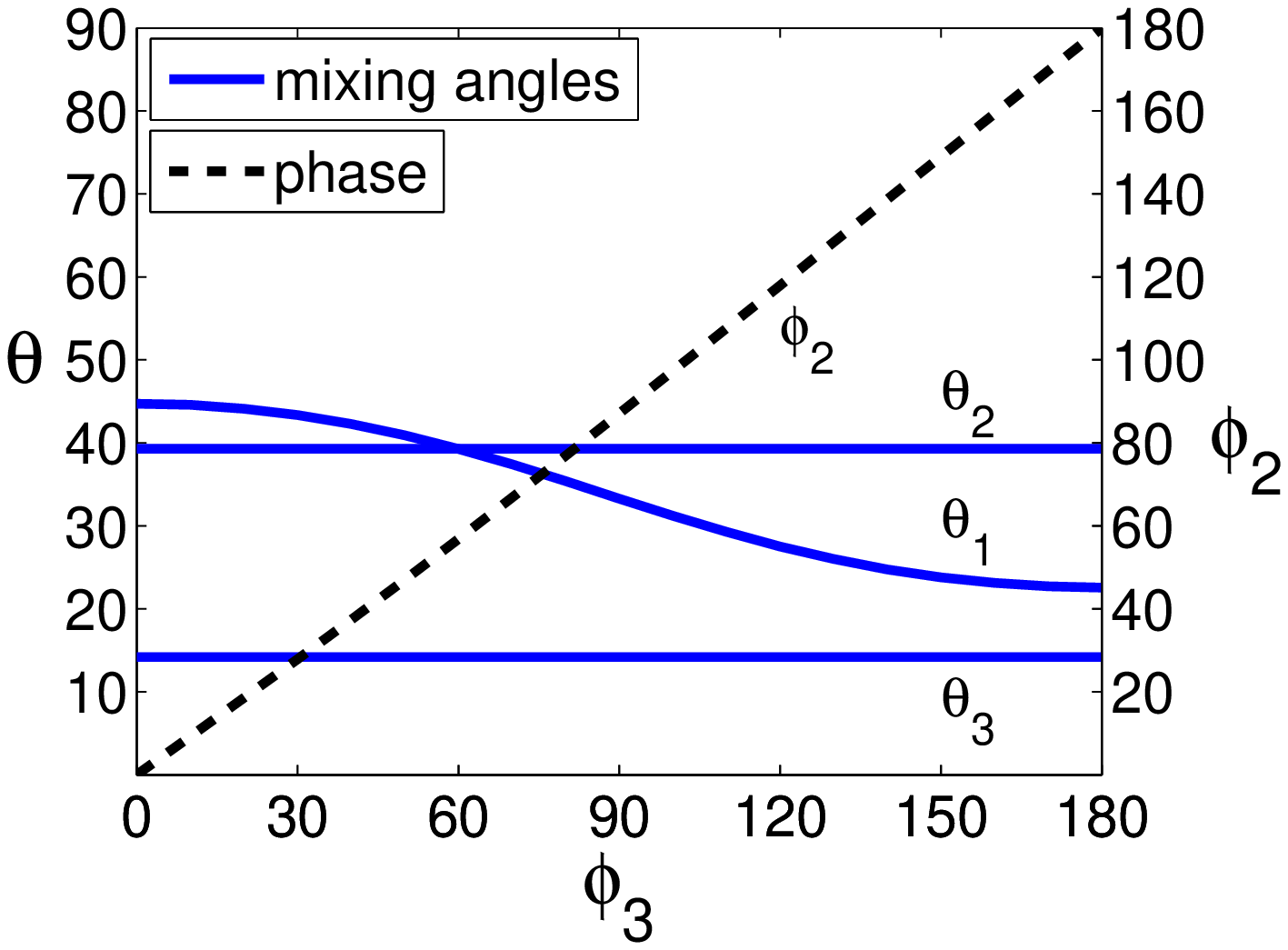}}
        \subfigure[~P3]{
          \includegraphics[width=5.8cm]{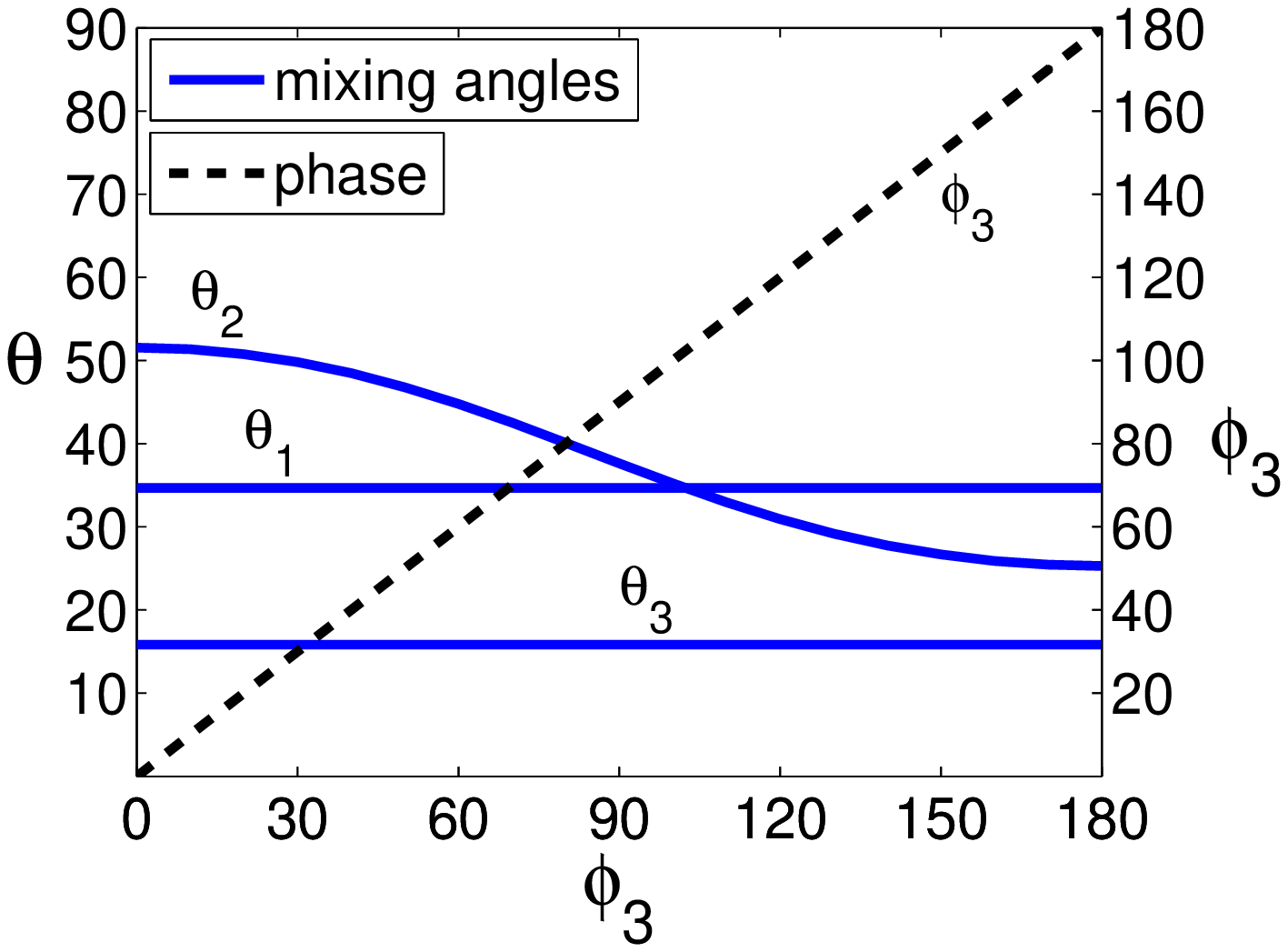}}
        \subfigure[~P4]{
          \includegraphics[width=5.8cm]{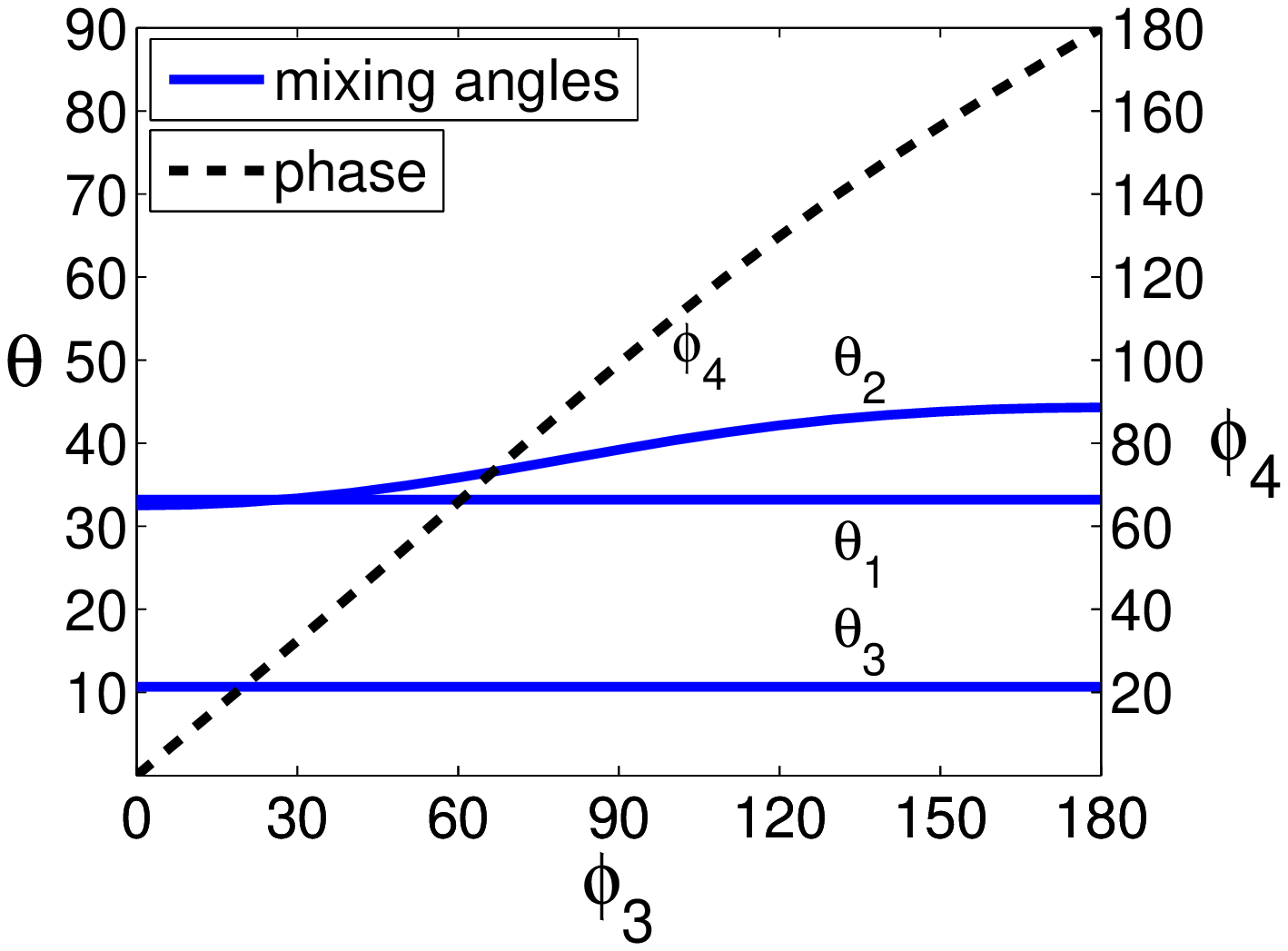}}
        \subfigure[~P5]{
          \includegraphics[width=5.8cm]{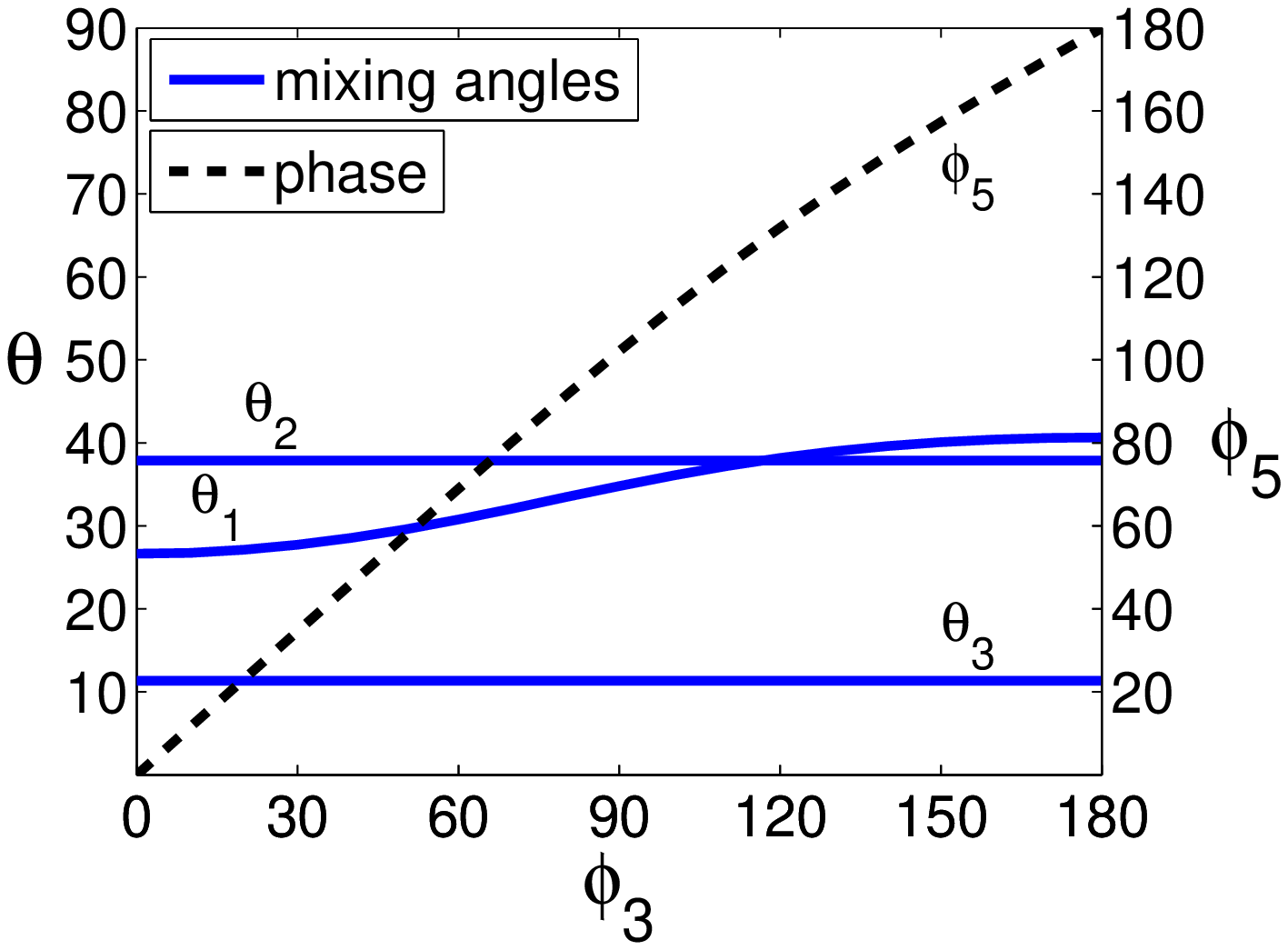}}
        \subfigure[~P6]{
          \includegraphics[width=5.8cm]{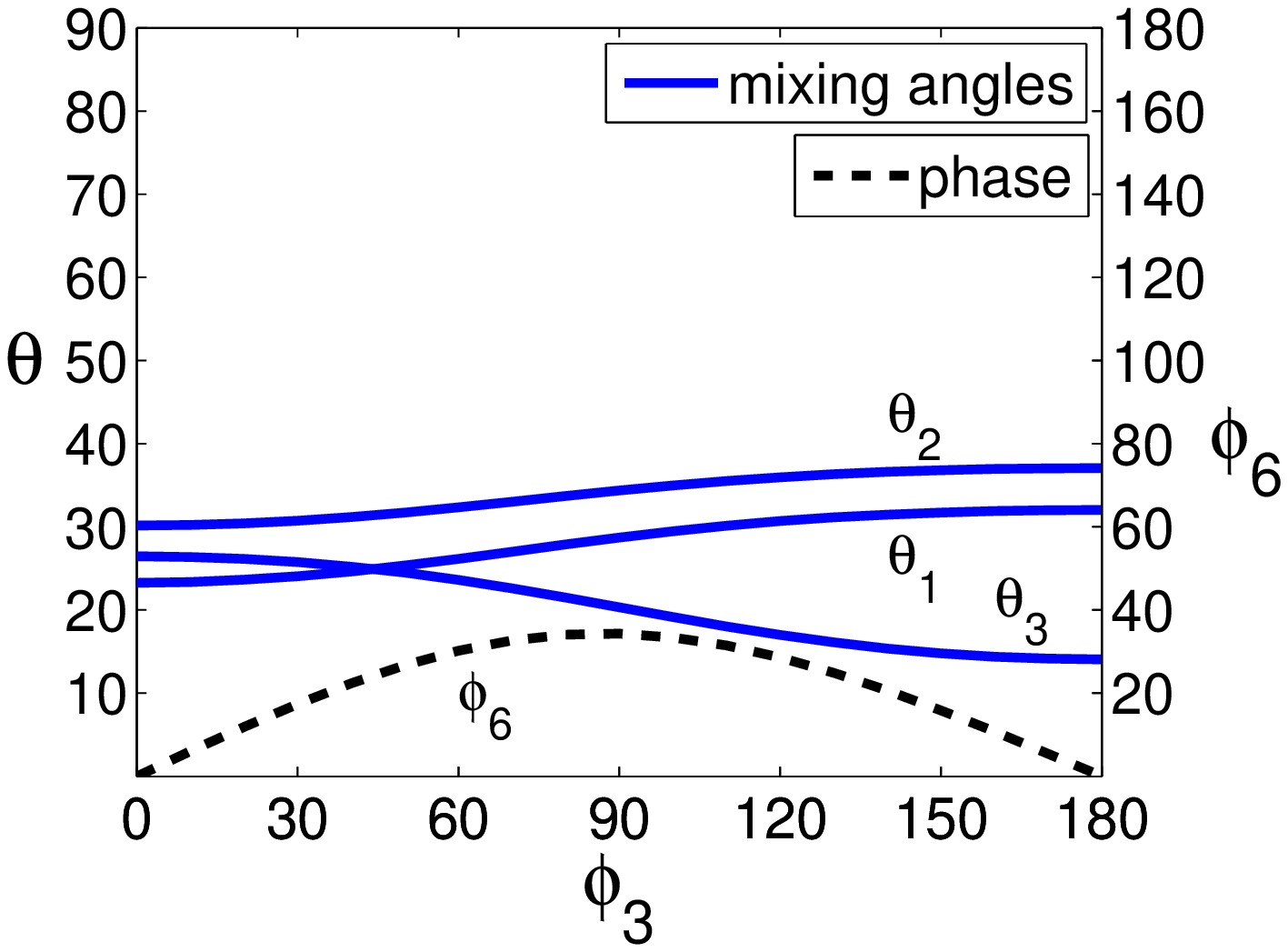}}
        \subfigure[~P7]{
          \includegraphics[width=5.8cm]{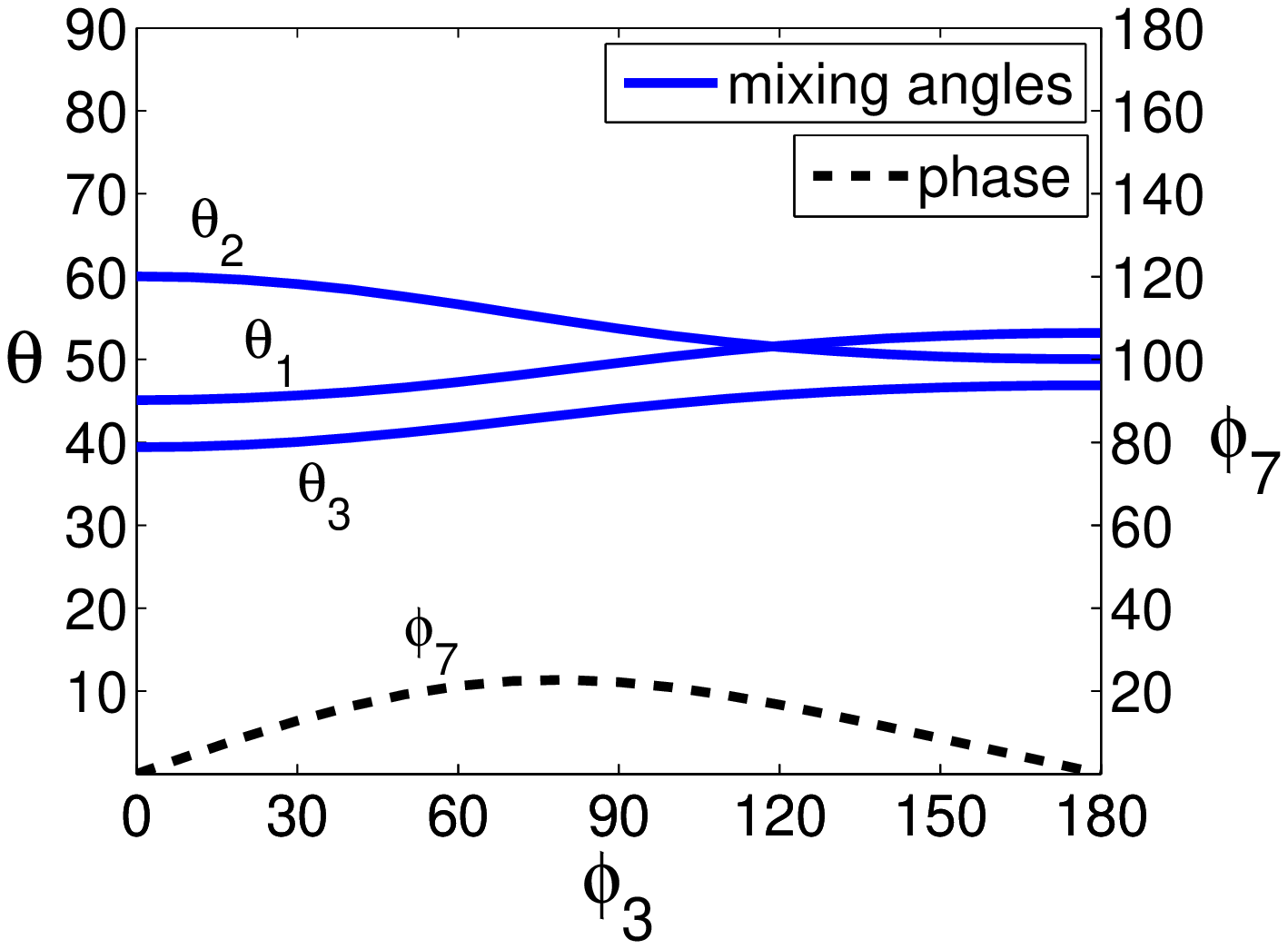}}
        \subfigure[~P8]{
          \includegraphics[width=5.8cm]{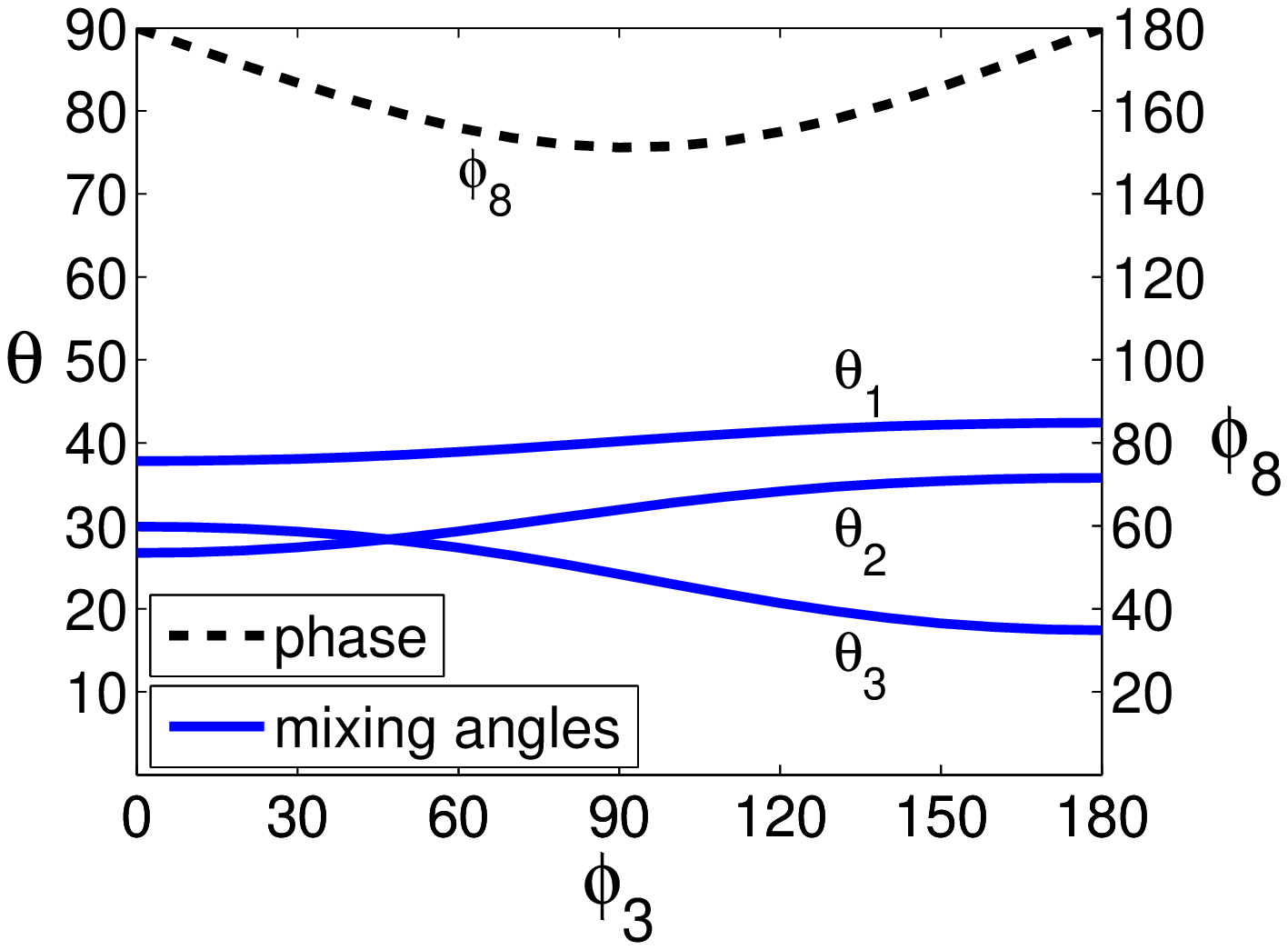}}
        \subfigure[~P9]{
          \includegraphics[width=5.8cm]{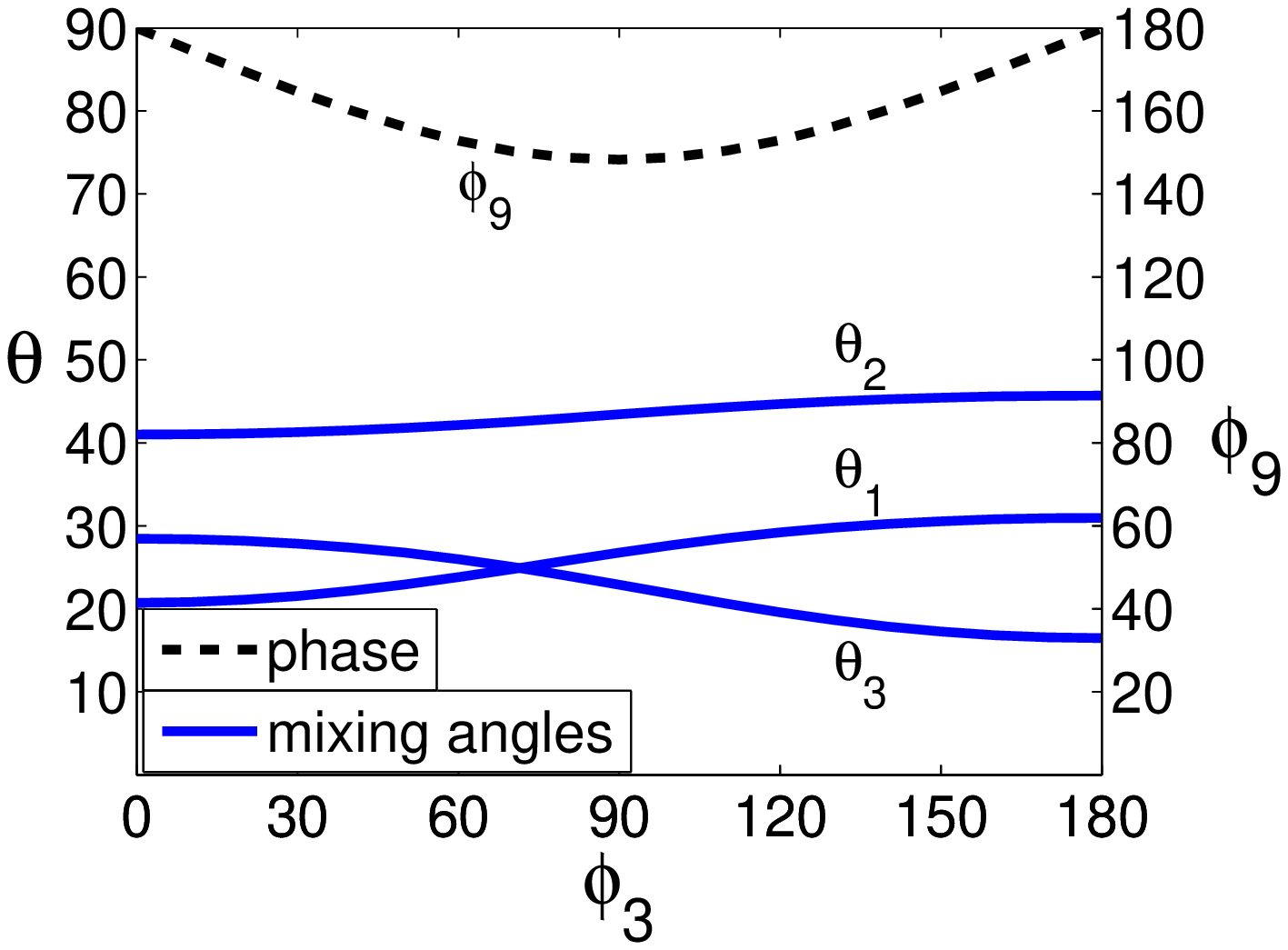}}
        \caption{Mixing angles and lepton {\it CP}-violating phases in different schemes (All the values are in the unit of degree~(${}^\circ$).)}
        \label{fig1}
      \end{figure*}

From the CKM and the PMNS matrices determined above, we then use
the formulas in Table~\ref{tabscheme} to determine all the four
parameters, i.e., three mixing angles and one {\it CP}-violating phase, in
the other schemes in both the quark and the lepton sectors. Then we
carry out calculations with other {\it CP}-violating phases by the same
procedure. The results are listed in Table~\ref{tab1}.
\begingroup
\squeezetable
\begin{table*}
      \caption{Mixing angles and {\it CP}-violating phases in different schemes}\label{tab1}
      \begin{ruledtabular}
\begin{scriptsize}
      \begin{tabular}{ccccccc}
        \toprule
        & quark & lepton($\phi_3=0^\circ$) & lepton($\phi_3=45^\circ$) & lepton($\phi_3=90^\circ$) & lepton($\phi_3=135^\circ$) & lepton($\phi_3=180^\circ$)\\
        \hline
        P1 & $\theta_{12}=\left(13.023^{+0.038}_{-0.038}\right)^\circ$ &$\vartheta_{12}=\left(33.65^{+1.11}_{-1.00}\right)^\circ$ &$\vartheta_{12}=\left(33.65^{+1.11}_{-1.00}\right)^\circ$ & $\vartheta_{12}=\left(33.65^{+1.11}_{-1.00}\right)^\circ$ &$\vartheta_{12}=\left(33.65^{+1.11}_{-1.00}\right)^\circ$ &$\vartheta_{12}=\left(33.65^{+1.11}_{-1.00}\right)^\circ$ \\
        & $\theta_{23}=\left(2.360^{+0.065}_{-0.038}\right)^\circ$ &$\vartheta_{23}=\left(38.41^{+1.40}_{-1.24}\right)^\circ$&$\vartheta_{23}=\left(38.41^{+1.40}_{-1.24}\right)^\circ$& $\vartheta_{23}=\left(38.41^{+1.40}_{-1.24}\right)^\circ$ &$\vartheta_{23}=\left(38.41^{+1.40}_{-1.24}\right)^\circ$&$\vartheta_{23}=\left(38.41^{+1.40}_{-1.24}\right)^\circ$\\
        & $\theta_{13}=\left(0.201^{+0.010}_{-0.008}\right)^\circ$ &$\vartheta_{13}=\left(8.93^{+0.46}_{-0.48}\right)^\circ$&$\vartheta_{13}=\left(8.93^{+0.46}_{-0.48}\right)^\circ$& $\vartheta_{13}=\left(8.93^{+0.46}_{-0.48}\right)^\circ$ &$\vartheta_{13}=\left(8.93^{+0.46}_{-0.48}\right)^\circ$&$\vartheta_{13}=\left(8.93^{+0.46}_{-0.48}\right)^\circ$\\
        & $\phi_1=\left(69.10^{+2.02}_{-3.85}\right)^\circ$ & $\phi_1=180^\circ$ & $\phi_1=\left(133.70^{+0.98}_{-0.88}\right)^\circ$ & $\phi_1=\left(83.37^{+4.96}_{-1.40}\right)^\circ$ & $\phi_1=\left(37.46^{+1.20}_{-1.11}\right)^\circ$& $\phi_1=0^\circ$ \\
      \\
        P2 & $\theta_1=\left(12.109^{+0.174}_{-0.327}\right)^\circ$ & $\vartheta_1=\left(44.73^{+1.33}_{-1.27}\right)^\circ$ & $\vartheta_1=\left(41.64^{+1.16}_{-1.08}\right)^\circ$ & $\vartheta_1=\left(33.28^{+1.10}_{-1.00}\right)^\circ$ & $\vartheta_1=\left(25.34^{+1.31}_{-1.20}\right)^\circ$ & $\vartheta_1=\left(22.57^{+1.36}_{-1.24}\right)^\circ$\\
        & $\theta_2=\left(2.369^{+0.065}_{-0.037}\right)^\circ$
         & $\vartheta_2=\left(39.28^{+1.37}_{-1.21}\right)^\circ$ & $\vartheta_2=\left(39.28^{+1.37}_{-1.21}\right)^\circ$ & $\vartheta_2=\left(39.28^{+1.37}_{-1.21}\right)^\circ$ & $\vartheta_2=\left(39.28^{+1.37}_{-1.21}\right)^\circ$ & $\vartheta_2=\left(39.28^{+1.37}_{-1.21}\right)^\circ$\\
        & $\theta_3=\left(4.880^{+0.256}_{-0.277}\right)^\circ$
         & $\vartheta_3=\left(14.19^{+0.80}_{-0.85}\right)^\circ$ & $\vartheta_3=\left(14.19^{+0.80}_{-0.85}\right)^\circ$ & $\vartheta_3=\left(14.19^{+0.80}_{-0.85}\right)^\circ$ & $\vartheta_3=\left(14.19^{+0.80}_{-0.85}\right)^\circ$ & $\vartheta_3=\left(14.19^{+0.80}_{-0.85}\right)^\circ$\\
        & $\phi_2=\left(89.69^{+2.29}_{-3.95}\right)^\circ$ & $\phi_2=0^\circ$ & $\phi_2=\left(42.20^{+0.99}_{-1.01}\right)^\circ$
         & $\phi_2=\left(87.07^{+1.05}_{-1.10}\right)^\circ$ & $\phi_2=\left(133.51^{+0.54}_{-0.57}\right)^\circ$& $\phi_2=180^\circ$ \\
      \\
        P3 & $\theta_1=\left(13.025^{+0.038}_{-0.038}\right)^\circ$ & $\vartheta_1=\left(34.68^{+1.07}_{-0.97}\right)^\circ$ & $\vartheta_1=\left(34.68^{+1.07}_{-0.97}\right)^\circ$ & $\vartheta_1=\left(34.68^{+1.07}_{-0.97}\right)^\circ$ & $\vartheta_1=\left(34.68^{+1.07}_{-0.97}\right)^\circ$ & $\vartheta_1=\left(34.68^{+1.07}_{-0.97}\right)^\circ$\\
        & $\theta_2=\left(2.205^{+0.068}_{-0.068}\right)^\circ$ & $\vartheta_2=\left(51.54^{+1.62}_{-1.51}\right)^\circ$ & $\vartheta_2=\left(47.67^{+1.60}_{-1.45}\right)^\circ$ & $\vartheta_2=\left(37.64^{+1.58}_{-1.40}\right)^\circ$ & $\vartheta_2=\left(28.44^{+1.64}_{-1.48}\right)^\circ$ & $\vartheta_2=\left(25.28^{+1.64}_{-1.48}\right)^\circ$\\
        & $\theta_3=\left(0.894^{+0.045}_{-0.045}\right)^\circ$ & $\vartheta_3=\left(15.83^{+0.88}_{-0.93}\right)^\circ$ & $\vartheta_3=\left(15.83^{+0.88}_{-0.93}\right)^\circ$ & $\vartheta_3=\left(15.83^{+0.88}_{-0.93}\right)^\circ$ & $\vartheta_3=\left(15.83^{+0.88}_{-0.93}\right)^\circ$ & $\vartheta_3=\left(15.83^{+0.88}_{-0.93}\right)^\circ$\\
        & $\phi_3=\left(89.29^{+3.99}_{-2.33}\right)^\circ$ & $\phi_3=0^\circ$ & $\phi_3=45.00^\circ$ & $\phi_3=90.00^\circ$ & $\phi_3=135.00^\circ$ & $\phi_3=180^\circ$ \\
      \\
        P4 & $\theta_1=\left(13.023^{+0.038}_{-0.038}\right)^\circ$ & $\vartheta_1=\left(33.19^{+1.09}_{-0.99}\right)^\circ$ & $\vartheta_1=\left(33.19^{+1.09}_{-0.99}\right)^\circ$ & $\vartheta_1=\left(33.19^{+1.09}_{-0.99}\right)^\circ$ & $\vartheta_1=\left(33.19^{+1.09}_{-0.99}\right)^\circ$ & $\vartheta_1=\left(33.19^{+1.09}_{-0.99}\right)^\circ$\\
        & $\theta_2=\left(2.377^{+0.066}_{-0.038}\right)^\circ$ & $\vartheta_2=\left(32.51^{+1.46}_{-1.30}\right)^\circ$ & $\vartheta_2=\left(34.43^{+1.34}_{-1.19}\right)^\circ$ & $\vartheta_2=\left(39.23^{+1.23}_{-1.09}\right)^\circ$ & $\vartheta_2=\left(43.13^{+1.37}_{-1.22}\right)^\circ$ & $\vartheta_2=\left(44.31^{+1.46}_{-1.30}\right)^\circ$\\
        & $\theta_3=\left(0.207^{+0.010}_{-0.010}\right)^\circ$ & $\vartheta_3=\left(10.69^{+0.56}_{-0.58}\right)^\circ$ & $\vartheta_3=\left(10.69^{+0.56}_{-0.58}\right)^\circ$ & $\vartheta_3=\left(10.69^{+0.56}_{-0.58}\right)^\circ$ & $\vartheta_3=\left(10.69^{+0.56}_{-0.58}\right)^\circ$ & $\vartheta_3=\left(10.69^{+0.56}_{-0.58}\right)^\circ$\\
        & $\phi_4=\left(111.95^{+3.82}_{-2.02}\right)^\circ$ & $\phi_4=0^\circ$ & $\phi_4=\left(49.01^{+1.36}_{-1.48}\right)^\circ$ & $\phi_4=\left(99.20^{+1.88}_{-2.08}\right)^\circ$ & $\phi_4=\left(143.60^{+1.33}_{-1.47}\right)^\circ$& $\phi_4=180^\circ$ \\
      \\
        P5 & $\theta_1=\left(13.026^{+0.038}_{-0.038}\right)^\circ$ & $\vartheta_1=\left(26.63^{+1.21}_{-1.12}\right)^\circ$ & $\vartheta_1=\left(29.03^{+1.15}_{-1.06}\right)^\circ$ & $\vartheta_1=\left(34.80^{+1.05}_{-0.95}\right)^\circ$ & $\vartheta_1=\left(39.32^{+1.14}_{-1.04}\right)^\circ$ & $\vartheta_1=\left(40.66^{+1.22}_{-1.11}\right)^\circ$\\
        & $\theta_2=\left(2.360^{+0.065}_{-0.038}\right)^\circ$ & $\vartheta_2=\left(37.86^{+1.38}_{-1.22}\right)^\circ$ & $\vartheta_2=\left(37.86^{+1.38}_{-1.22}\right)^\circ$ & $\vartheta_2=\left(37.86^{+1.38}_{-1.22}\right)^\circ$ & $\vartheta_2=\left(37.86^{+1.38}_{-1.22}\right)^\circ$ & $\vartheta_2=\left(37.86^{+1.38}_{-1.22}\right)^\circ$\\
        & $\theta_3=\left(0.202^{+0.010}_{-0.010}\right)^\circ$ & $\vartheta_3=\left(11.34^{+0.61}_{-0.63}\right)^\circ$ & $\vartheta_3=\left(11.34^{+0.61}_{-0.63}\right)^\circ$ & $\vartheta_3=\left(11.34^{+0.61}_{-0.63}\right)^\circ$ & $\vartheta_3=\left(11.34^{+0.61}_{-0.63}\right)^\circ$ & $\vartheta_3=\left(11.34^{+0.61}_{-0.63}\right)^\circ$\\
        & $\phi_5=\left(110.94^{+3.85}_{-2.02}\right)^\circ$ & $\phi_5=0^\circ$ & $\phi_5=\left(51.82^{+0.80}_{-0.86}\right)^\circ$ & $\phi_5=\left(102.03^{+1.46}_{-1.57}\right)^\circ$ & $\phi_5=\left(145.09^{+1.17}_{-1.26}\right)^\circ$& $\phi_5=180^\circ$ \\
      \\
        P6 & $\theta_1=\left(13.016^{+0.038}_{-0.038}\right)^\circ$ & $\vartheta_1=\left(23.28^{+1.23}_{-1.18}\right)^\circ$ & $\vartheta_1=\left(24.98^{+1.08}_{-1.13}\right)^\circ$ & $\vartheta_1=\left(28.72^{+1.05}_{-1.00}\right)^\circ$ & $\vartheta_1=\left(31.31^{+0.99}_{-0.92}\right)^\circ$ & $\vartheta_1=\left(32.03^{+0.98}_{-0.91}\right)^\circ$\\
        & $\theta_2=\left(2.316^{+0.064}_{-0.037}\right)^\circ$ & $\vartheta_2=\left(30.16^{+1.52}_{-1.37}\right)^\circ$ & $\vartheta_2=\left(31.44^{+1.41}_{-1.26}\right)^\circ$ & $\vartheta_2=\left(34.36^{+1.27}_{-1.13}\right)^\circ$ & $\vartheta_2=\left(36.47^{+1.25}_{-1.12}\right)^\circ$ & $\vartheta_2=\left(37.06^{+1.26}_{-1.13}\right)^\circ$\\
        & $\theta_3=\left(0.497^{+0.015}_{-0.015}\right)^\circ$ & $\vartheta_3=\left(26.46^{+0.85}_{-0.79}\right)^\circ$ & $\vartheta_3=\left(24.87^{+0.87}_{-0.80}\right)^\circ$ & $\vartheta_3=\left(20.33^{+0.96}_{-0.87}\right)^\circ$ & $\vartheta_3=\left(15.72^{+1.04}_{-0.94}\right)^\circ$ & $\vartheta_3=\left(14.06^{+1.05}_{-0.95}\right)^\circ$\\
        & $\phi_6=\left(22.72^{+1.25}_{-1.18}\right)^\circ$ & $\phi_6=0^\circ$ & $\phi_6=\left(24.65^{+1.40}_{-1.47}\right)^\circ$ & $\phi_6=\left(34.29^{+1.92}_{-2.02}\right)^\circ$ & $\phi_6=\left(22.85^{+1.16}_{-1.23}\right)^\circ$& $\phi_6=0^\circ$ \\
      \\
        P7 & $\theta_1=\left(13.235^{+0.039}_{-0.038}\right)^\circ$ & $\vartheta_1=\left(45.11^{+1.09}_{-0.95}\right)^\circ$ & $\vartheta_1=\left(46.35^{+1.06}_{-0.93}\right)^\circ$ & $\vartheta_1=\left(49.59^{+1.06}_{-0.93}\right)^\circ$ & $\vartheta_1=\left(52.36^{+1.20}_{-1.06}\right)^\circ$ & $\vartheta_1=\left(53.21^{+1.27}_{-1.12}\right)^\circ$\\
        & $\theta_2=\left(10.363^{+0.283}_{-0.164}\right)^\circ$ & $\vartheta_2=\left(60.03^{+1.76}_{-1.69}\right)^\circ$ & $\vartheta_2=\left(58.03^{+1.71}_{-1.63}\right)^\circ$ & $\vartheta_2=\left(53.72^{+1.56}_{-1.46}\right)^\circ$ & $\vartheta_2=\left(50.82^{+1.40}_{-1.31}\right)^\circ$ & $\vartheta_2=\left(50.03^{+1.35}_{-1.27}\right)^\circ$\\
        & $\theta_3=\left(10.167^{+0.276}_{-0.160}\right)^\circ$ & $\vartheta_3=\left(39.41^{+1.64}_{-1.67}\right)^\circ$ & $\vartheta_3=\left(40.84^{+1.50}_{-1.54}\right)^\circ$ & $\vartheta_3=\left(44.04^{+1.31}_{-1.36}\right)^\circ$ & $\vartheta_3=\left(46.27^{+1.23}_{-1.28}\right)^\circ$ & $\vartheta_3=\left(46.88^{+1.21}_{-1.25}\right)^\circ$\\
        & $\phi_7=\left(1.08^{+0.06}_{-0.06}\right)^\circ$ & $\phi_7=0^\circ$ & $\phi_7=\left(17.84^{+0.96}_{-1.00}\right)^\circ$ & $\phi_7=\left(22.16^{+1.18}_{-1.20}\right)^\circ$ & $\phi_7=\left(12.76^{+0.71}_{-0.68}\right)^\circ$& $\phi_7=0^\circ$ \\
      \\
        P8 & $\theta_1=\left(13.034^{+0.038}_{-0.038}\right)^\circ$ & $\vartheta_1=\left(37.80^{+1.17}_{-1.05}\right)^\circ$ & $\vartheta_1=\left(38.41^{+1.20}_{-1.08}\right)^\circ$ & $\vartheta_1=\left(40.17^{+1.27}_{-1.15}\right)^\circ$ & $\vartheta_1=\left(41.87^{+1.41}_{-1.27}\right)^\circ$ & $\vartheta_1=\left(42.43^{+1.48}_{-1.32}\right)^\circ$\\
        & $\theta_2=\left(2.316^{+0.064}_{-0.037}\right)^\circ$ & $\vartheta_2=\left(26.73^{+1.23}_{-1.15}\right)^\circ$ & $\vartheta_2=\left(28.24^{+1.13}_{-1.05}\right)^\circ$ & $\vartheta_2=\left(31.95^{+1.02}_{-0.94}\right)^\circ$ & $\vartheta_2=\left(34.90^{+1.07}_{-0.99}\right)^\circ$ & $\vartheta_2=\left(35.77^{+1.10}_{-1.01}\right)^\circ$\\
        & $\theta_3=\left(0.497^{+0.015}_{-0.015}\right)^\circ$ & $\vartheta_3=\left(29.92^{+1.13}_{-1.01}\right)^\circ$ & $\vartheta_3=\left(28.52^{+1.19}_{-1.05}\right)^\circ$ & $\vartheta_3=\left(24.17^{+1.33}_{-1.18}\right)^\circ$ & $\vartheta_3=\left(19.29^{+1.44}_{-1.27}\right)^\circ$ & $\vartheta_3=\left(17.43^{+1.45}_{-1.28}\right)^\circ$\\
        & $\phi_8=\left(157.31^{+1.18}_{-1.25}\right)^\circ$ & $\phi_8=180^\circ$ & $\phi_8=\left(160.86^{+1.27}_{-1.20}\right)^\circ$ & $\phi_8=\left(151.21^{+1.79}_{-1.70}\right)^\circ$ & $\phi_8=\left(159.70^{+1.14}_{-1.08}\right)^\circ$& $\phi_8=180^\circ$ \\
      \\
        P9 & $\theta_1=\left(13.015^{+0.038}_{-0.038}\right)^\circ$ & $\vartheta_1=\left(20.72^{+1.05}_{-1.04}\right)^\circ$ & $\vartheta_1=\left(22.53^{+1.02}_{-1.02}\right)^\circ$ & $\vartheta_1=\left(26.78^{+0.93}_{-0.93}\right)^\circ$ & $\vartheta_1=\left(30.02^{+0.89}_{-0.87}\right)^\circ$ & $\vartheta_1=\left(30.96^{+0.89}_{-0.86}\right)^\circ$\\
        & $\theta_2=\left(2.423^{+0.067}_{-0.038}\right)^\circ$ & $\vartheta_2=\left(41.01^{+1.35}_{-1.21}\right)^\circ$ & $\vartheta_2=\left(41.64^{+1.36}_{-1.21}\right)^\circ$ & $\vartheta_2=\left(43.43^{+1.27}_{-1.12}\right)^\circ$ & $\vartheta_2=\left(45.14^{+1.57}_{-1.40}\right)^\circ$ & $\vartheta_2=\left(45.70^{+1.64}_{-1.46}\right)^\circ$\\
        & $\theta_3=\left(0.510^{+0.016}_{-0.016}\right)^\circ$ & $\vartheta_3=\left(28.45^{+0.96}_{-0.88}\right)^\circ$ & $\vartheta_3=\left(27.09^{+0.99}_{-0.90}\right)^\circ$ & $\vartheta_3=\left(22.90^{+1.12}_{-1.01}\right)^\circ$ & $\vartheta_3=\left(18.24^{+1.25}_{-1.13}\right)^\circ$ & $\vartheta_3=\left(16.46^{+2.70}_{-1.38}\right)^\circ$\\
        & $\phi_9=\left(158.32^{+1.13}_{-1.20}\right)^\circ$ & $\phi_9=180^\circ$ & $\phi_9=\left(158.05^{+1.24}_{-1.17}\right)^\circ$ & $\phi_9=\left(148.28^{+1.87}_{-1.77}\right)^\circ$ & $\phi_9=\left(158.21^{+1.27}_{-1.14}\right)^\circ$& $\phi_9=180^\circ$ \\
      \bottomrule
      \end{tabular}
\end{scriptsize}
      \end{ruledtabular}
\end{table*}
\endgroup
To better illustrate the dependence of the lepton mixing angles and
{\it CP}-violating phases on $\phi_3$, we also draw
a series of graphs in Fig.~\ref{fig1}.
From Fig.~\ref{fig1}, we can easily see that $\vartheta_1$ in P2,
$\vartheta_2$ in P3, $\vartheta_2$ in P4, $\vartheta_1$ in P5, and
all the mixing angles in P6$\sim$P9 have different values when
different {\it CP}-violating phases in the KM scheme are adopted.

It is necessary to explain here the assumptions used in our calculations. Since
generally there is just a slight dependence of the results of lepton
mixing angles in the CK scheme on the lepton {\it CP}-violating phases, we simply
assume that these results (\ref{normal}) are independent of the lepton
{\it CP}-violating phases. Actually, for some experimental groups, their results and
error bars of mixing angles in the CK scheme actually vary with different phase
assumptions.~(See for example Ref.~\cite{T2K} and Ref.~\cite{MINOS}.) This
proves that the independence is suitable only approximately.

\section{\label{sec3}QUARK-LEPTON COMPLEMENTARITY}
\begin{figure*}
        \centering
        \subfigure[~P1]{
          \includegraphics[width=5.8cm]{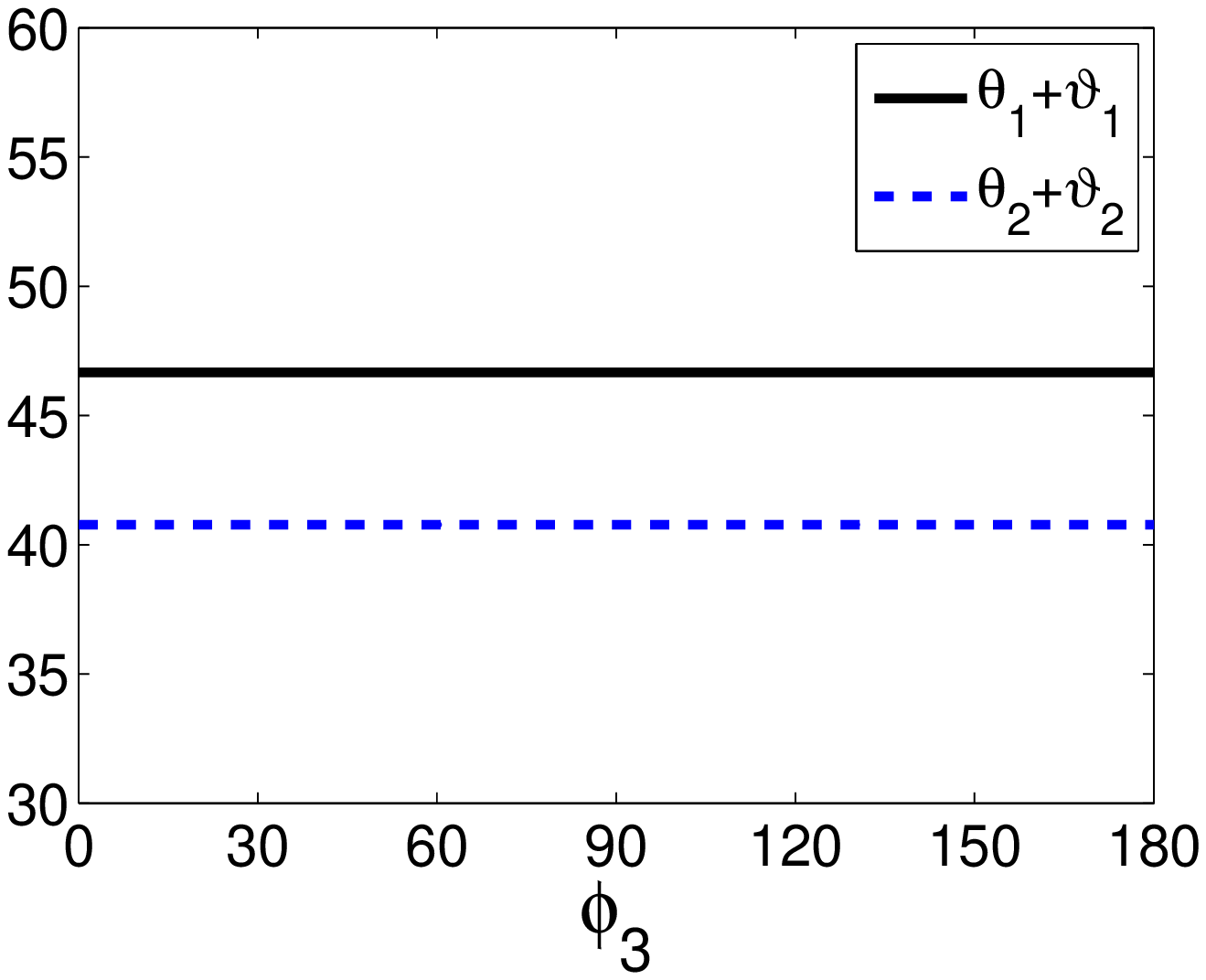}}
        \subfigure[~P2]{
          \includegraphics[width=5.8cm]{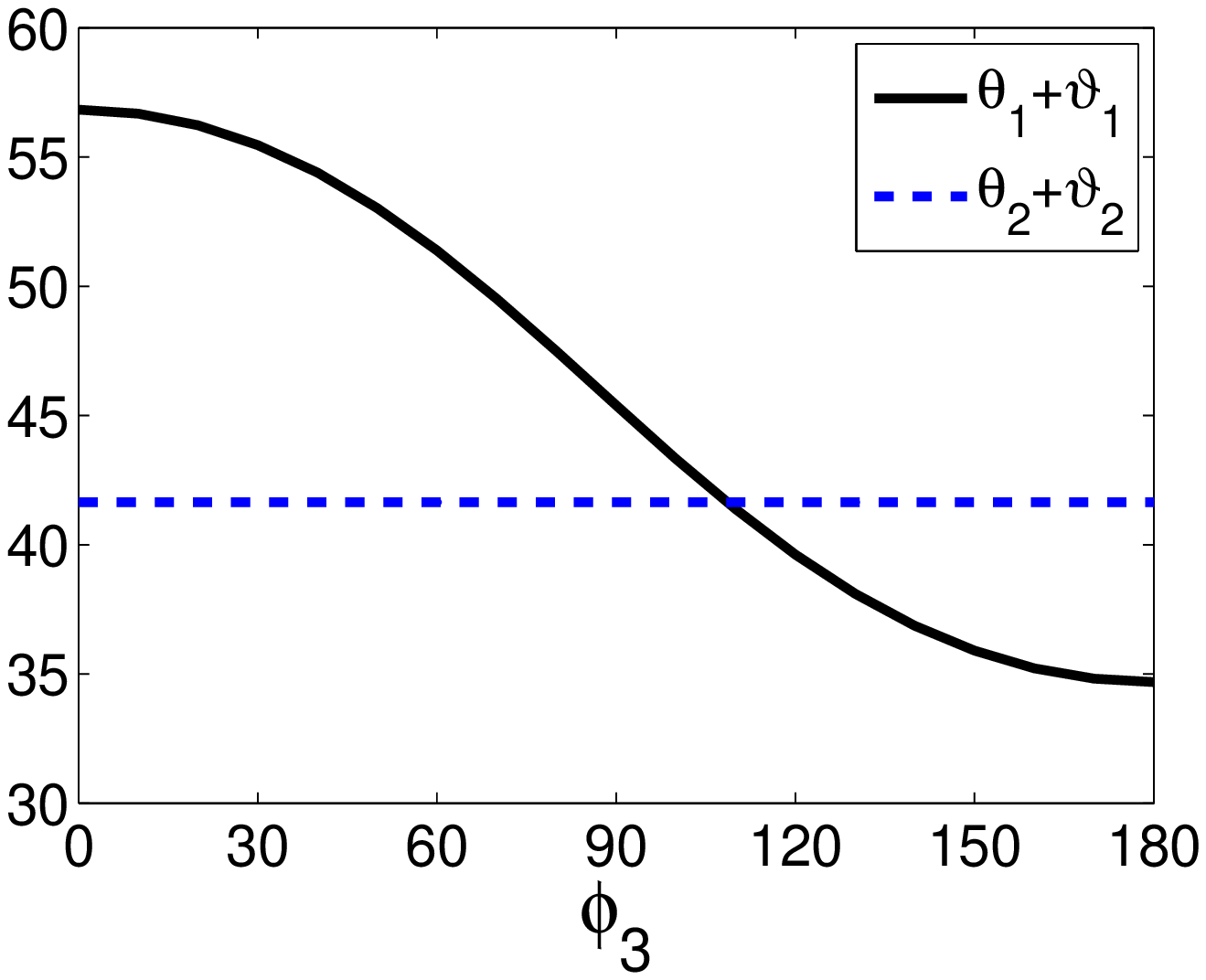}}
        \subfigure[~P3]{
          \includegraphics[width=5.8cm]{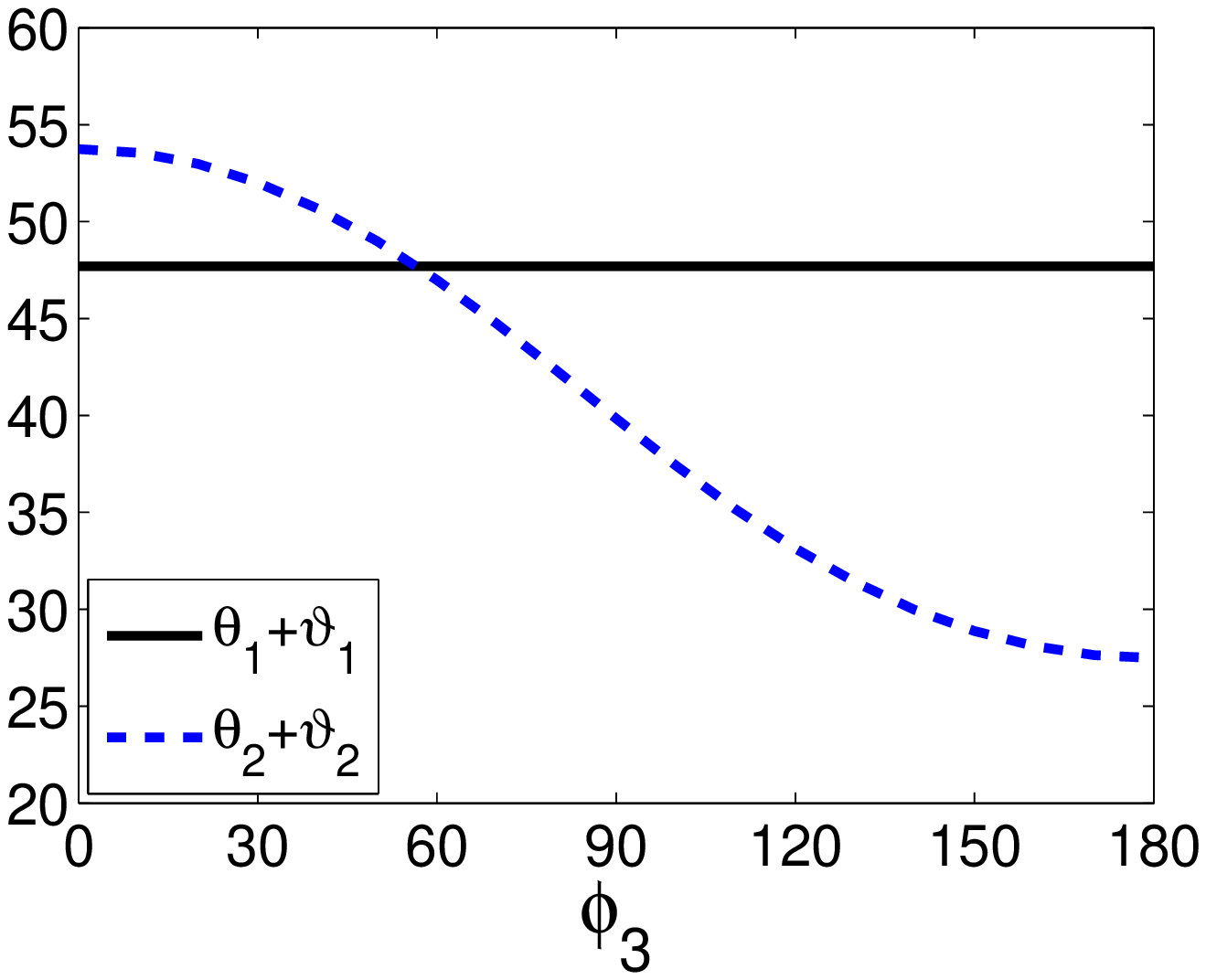}}
        \subfigure[~P4]{
          \includegraphics[width=5.8cm]{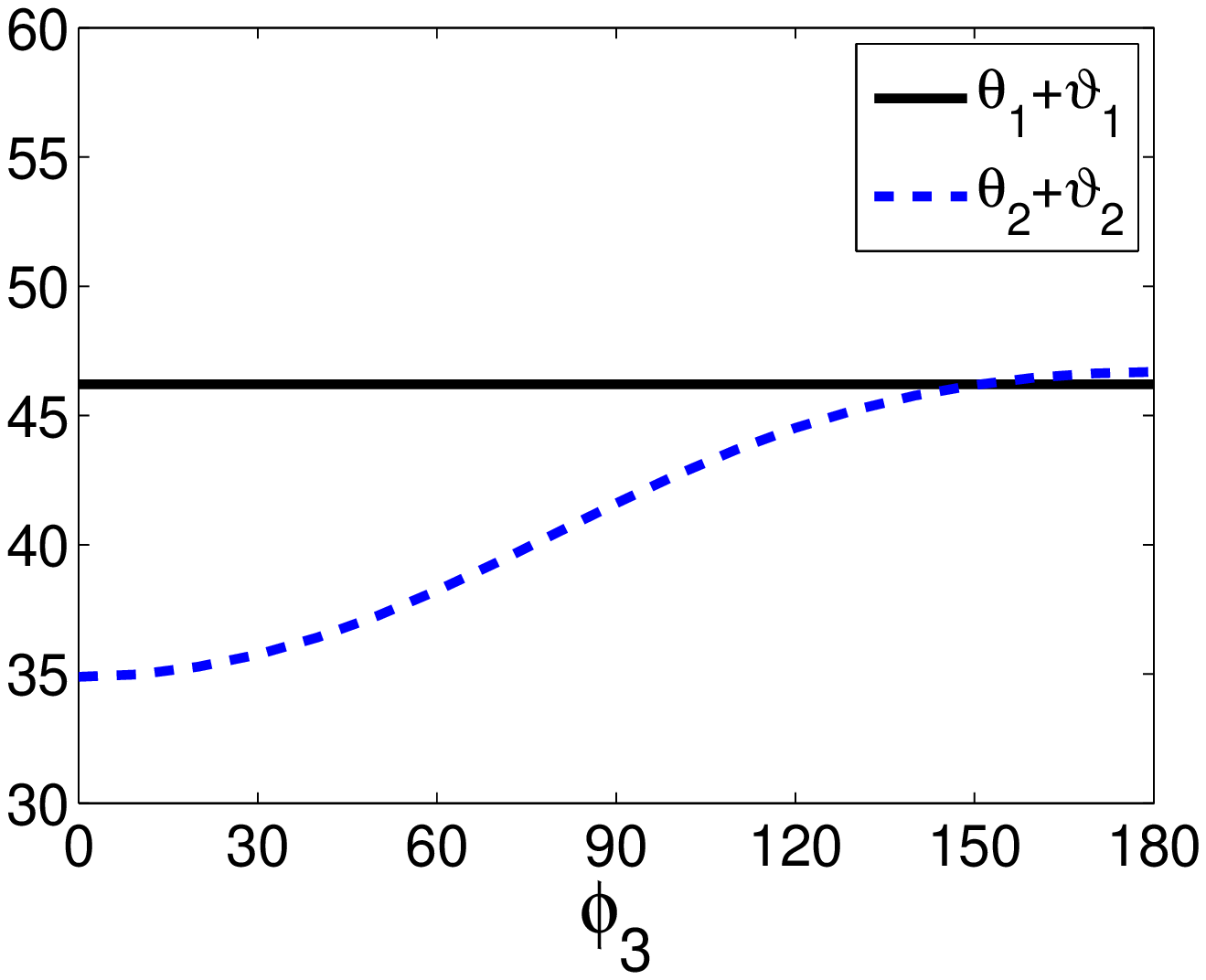}}
        \subfigure[~P5]{
          \includegraphics[width=5.8cm]{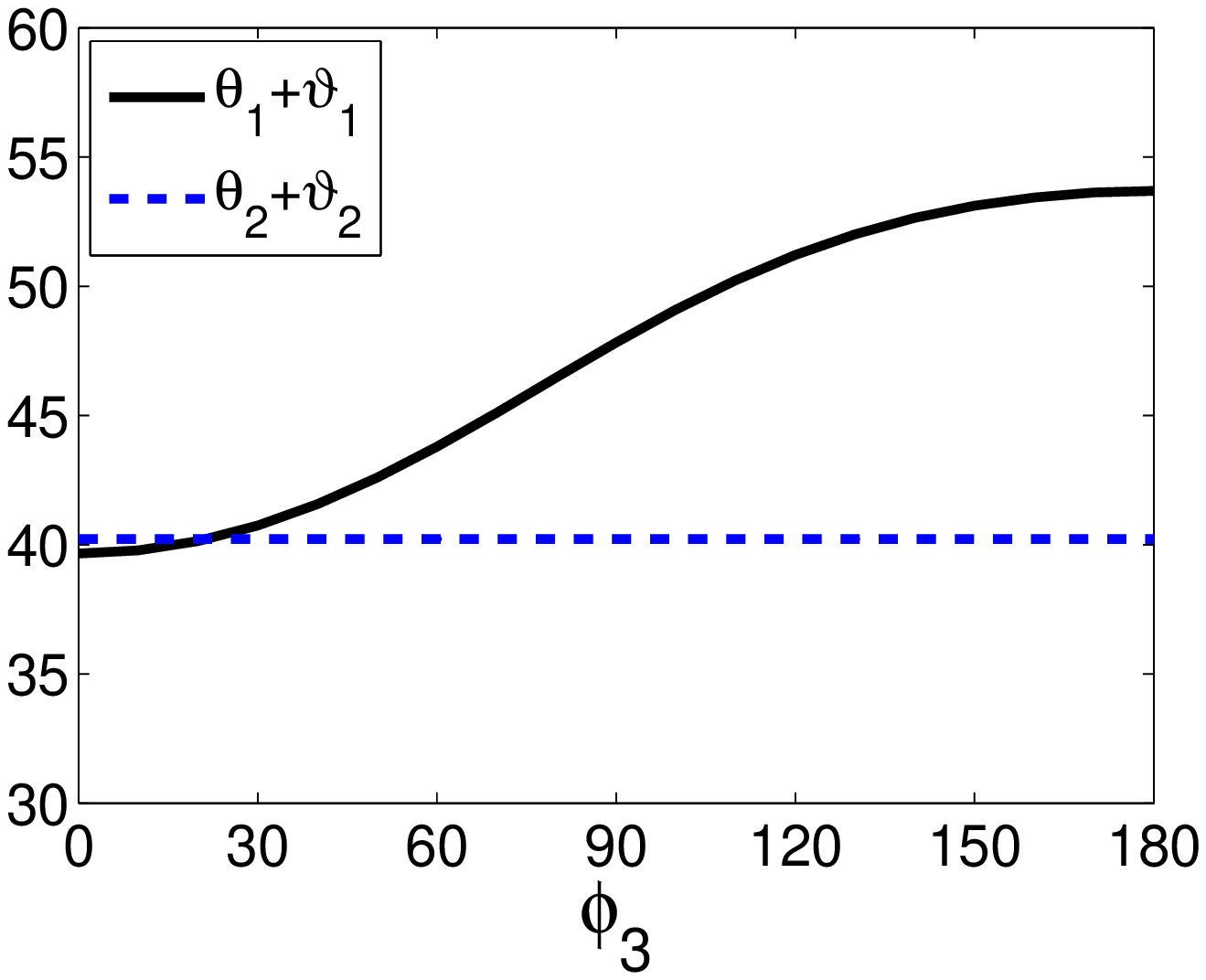}}
        \subfigure[~P6]{
          \includegraphics[width=5.8cm]{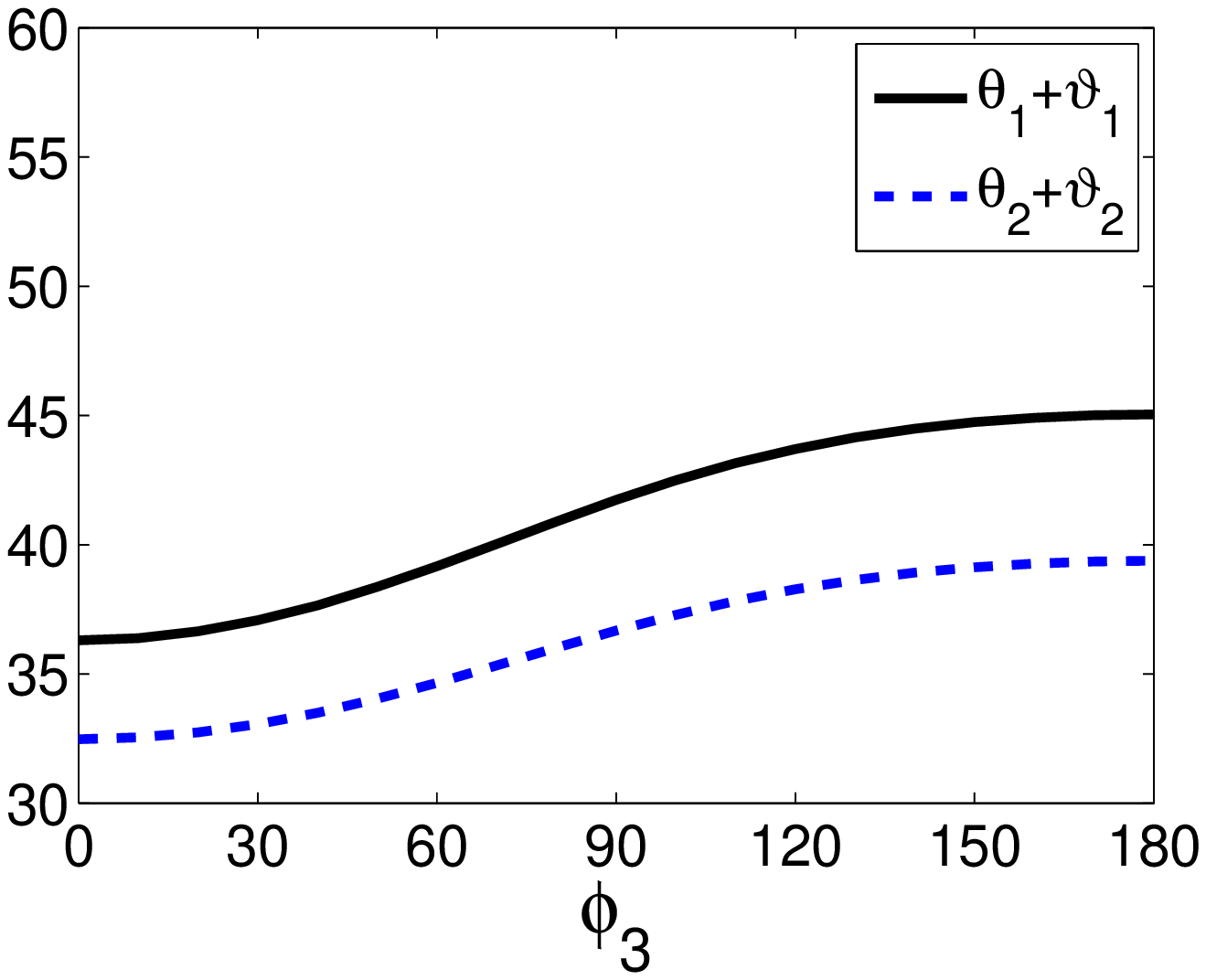}}
        \subfigure[~P7]{
          \includegraphics[width=5.8cm]{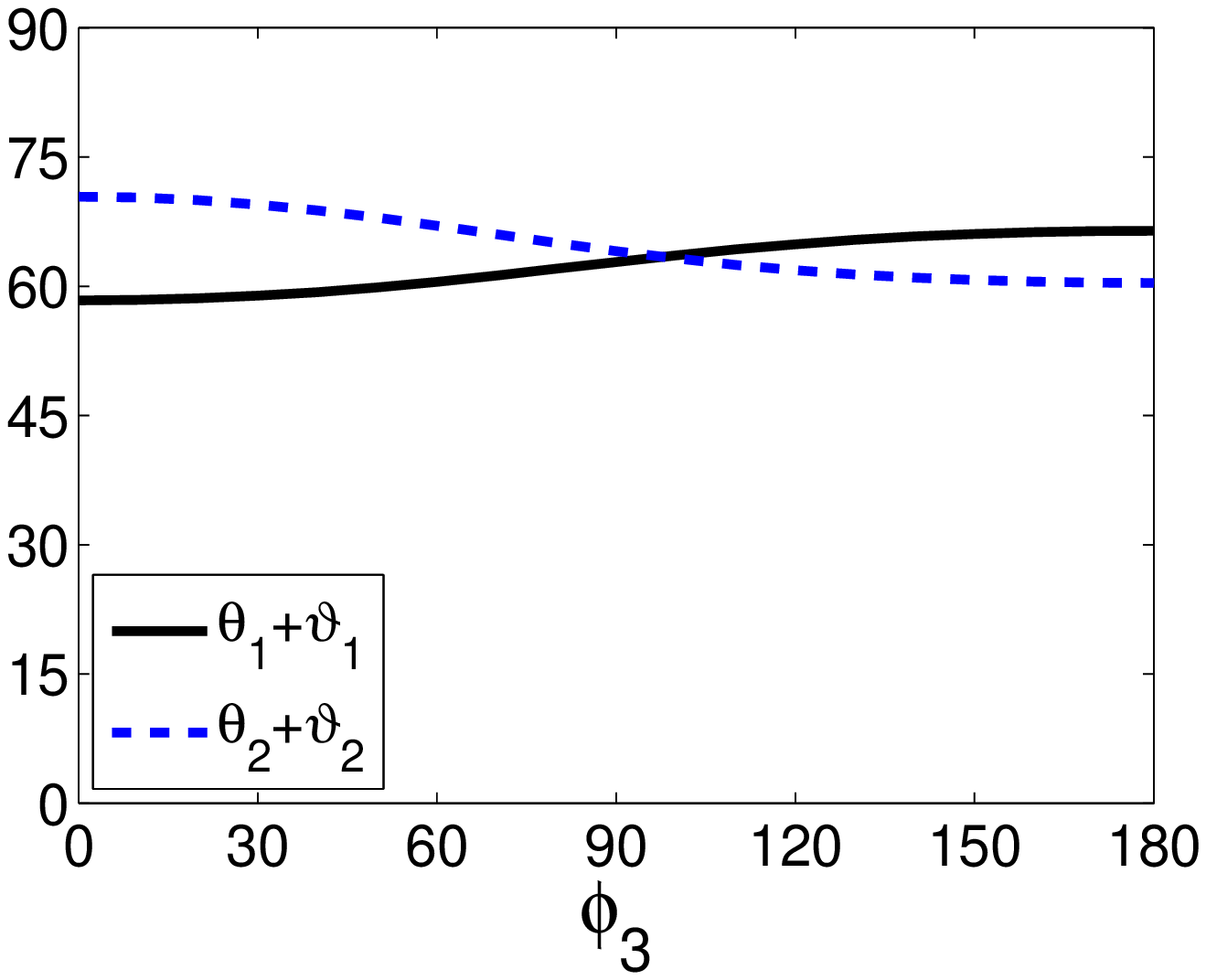}}
        \subfigure[~P8]{
          \includegraphics[width=5.8cm]{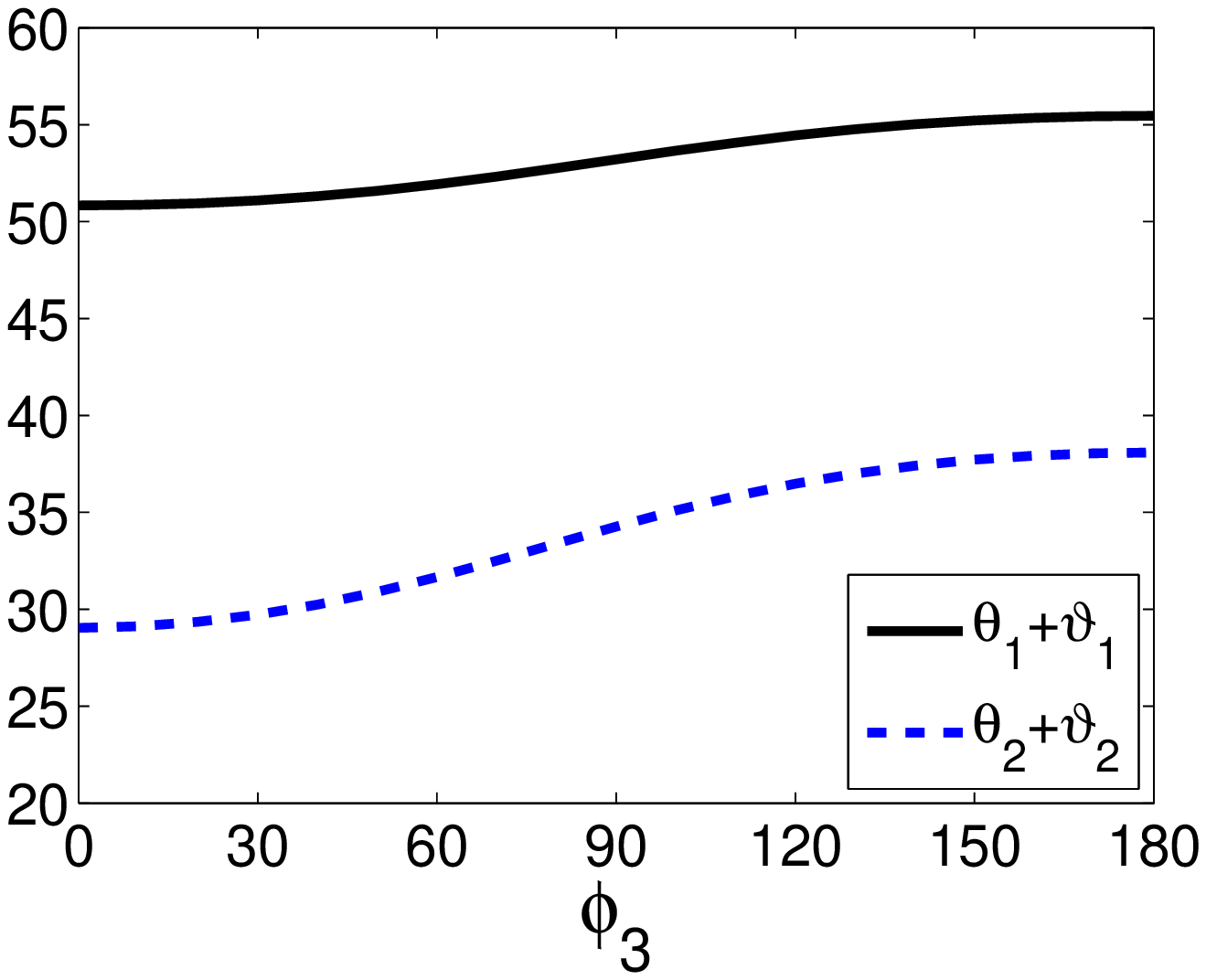}}
        \subfigure[~P9]{
          \includegraphics[width=5.8cm]{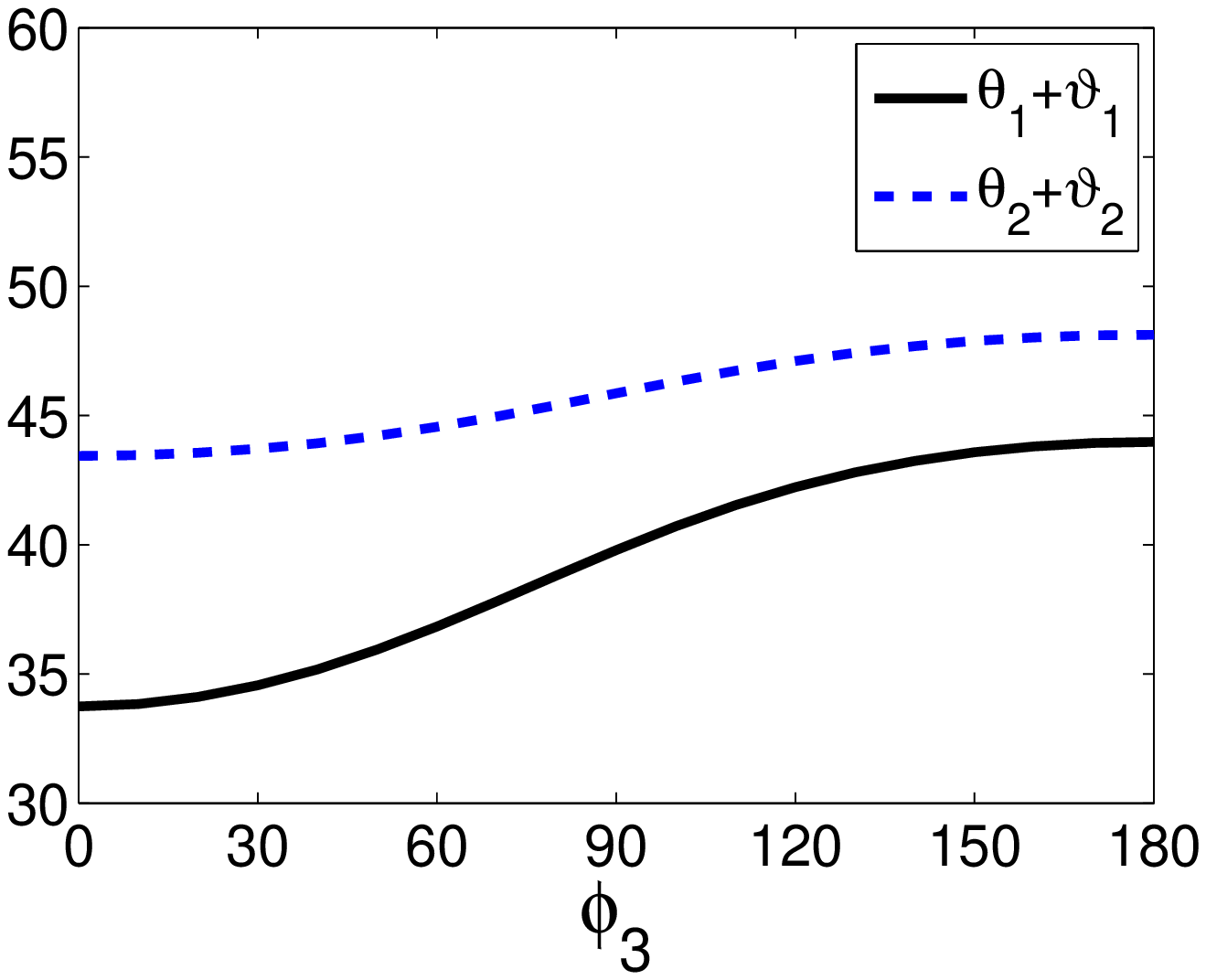}}
        \caption{The quark-lepton complementarity between mixing angles of quarks and leptons [All the values are in the unit of degree~(${}^\circ$).]}
        \label{fig3}
    \end{figure*}

With the quark and lepton mixing angles calculated in the previous section, we now go on to discuss the topic of QLC. A diagrammatic presentation of QLC is shown in Fig.~\ref{fig3}.
We list the results in Table~\ref{complementarity} with five different {\it CP}-violating phases, i.e., $\phi_3=0^\circ$, $45^\circ$, $90^\circ$, $135^\circ$, and $180^\circ$, respectively.
\begingroup
\squeezetable
\begin{table*}
      \caption{The quark-lepton complementarity between quark and lepton mixing angles}\label{complementarity}
      \begin{ruledtabular}
      \begin{tabular}{cccccccc}
        \toprule
        & QLC & $(\phi_3=0^\circ)$ & $(\phi_3=45^\circ)$ & $(\phi_3=90^\circ)$ & $(\phi_3=135^\circ)$ & $(\phi_3=180^\circ)$ & Type\\
        \hline
        P1 & $\theta_{12}+\vartheta_{12}$ & $\left(46.65^{+1.11}_{-1.00}\right)^\circ$ & $\left(46.65^{+1.11}_{-1.00}\right)^\circ$ & $\left(46.65^{+1.11}_{-1.00}\right)^\circ$ & $\left(46.65^{+1.11}_{-1.00}\right)^\circ$ & $\left(46.65^{+1.11}_{-1.00}\right)^\circ$ & $A_{1}$\\
        & $\theta_{23}+\vartheta_{23}$ & $\left(40.77^{+1.40}_{-1.24}\right)^\circ$ & $\left(40.77^{+1.40}_{-1.24}\right)^\circ$ & $\left(40.77^{+1.40}_{-1.24}\right)^\circ$ & $\left(40.77^{+1.40}_{-1.24}\right)^\circ$ & $\left(40.77^{+1.40}_{-1.24}\right)^\circ$ & $A_{4}$\\
        \hline
        P2 & $\theta_{1}+\vartheta_{1}$ & $\left(56.84^{+1.33}_{-1.27}\right)^\circ$ & $\left(53.75^{+1.16}_{-1.08}\right)^\circ$ & $\left(45.39^{+1.10}_{-1.00}\right)^\circ$ & $\left(37.45^{+1.31}_{-1.20}\right)^\circ$ & $\left(34.68^{+1.36}_{-1.24}\right)^\circ$ & $B_{1}$\\
        & $\theta_{2}+\vartheta_{2}$ & $\left(41.65^{+1.33}_{-1.27}\right)^\circ$ & $\left(41.65^{+1.33}_{-1.27}\right)^\circ$ & $\left(41.65^{+1.33}_{-1.27}\right)^\circ$ & $\left(41.65^{+1.33}_{-1.27}\right)^\circ$ & $\left(41.65^{+1.33}_{-1.27}\right)^\circ$ & $A_{3}$\\
        \hline
        P3 & $\theta_{1}+\vartheta_{1}$ & $\left(47.71^{+1.07}_{-0.97}\right)^\circ$ & $\left(47.71^{+1.07}_{-0.97}\right)^\circ$ & $\left(47.71^{+1.07}_{-0.97}\right)^\circ$ & $\left(47.71^{+1.07}_{-0.97}\right)^\circ$ & $\left(47.71^{+1.07}_{-0.97}\right)^\circ$ & $A_{3}$\\
        & $\theta_{2}+\vartheta_{2}$ & $\left(53.75^{+1.62}_{-1.51}\right)^\circ$ & $\left(49.88^{+1.60}_{-1.45}\right)^\circ$ & $\left(39.85^{+1.58}_{-1.40}\right)^\circ$ & $\left(30.65^{+1.64}_{-1.48}\right)^\circ$ & $\left(27.49^{+1.64}_{-1.48}\right)^\circ$ & $B_{4}$\\
        \hline
        P4 & $\theta_{1}+\vartheta_{1}$ & $\left(46.21^{+1.09}_{-0.99}\right)^\circ$ & $\left(46.21^{+1.09}_{-0.99}\right)^\circ$ & $\left(46.21^{+1.09}_{-0.99}\right)^\circ$ & $\left(46.21^{+1.09}_{-0.99}\right)^\circ$ & $\left(46.21^{+1.09}_{-0.99}\right)^\circ$ & $A_{2}$\\
        & $\theta_{2}+\vartheta_{2}$ & $\left(34.89^{+1.46}_{-1.30}\right)^\circ$ & $\left(36.81^{+1.34}_{-1.19}\right)^\circ$ & $\left(41.61^{+1.23}_{-1.09}\right)^\circ$ & $\left(45.51^{+1.37}_{-1.22}\right)^\circ$ & $\left(46.69^{+1.46}_{-1.30}\right)^\circ$ & $B_{4}$\\
        \hline
        P5 & $\theta_{1}+\vartheta_{1}$ & $\left(39.64^{+1.21}_{-1.12}\right)^\circ$ & $\left(42.06^{+1.15}_{-1.06}\right)^\circ$ & $\left(47.83^{+1.05}_{-0.95}\right)^\circ$ & $\left(52.35^{+1.14}_{-1.04}\right)^\circ$ & $\left(53.69^{+1.22}_{-1.11}\right)^\circ$ & $B_{3}$\\
        & $\theta_{2}+\vartheta_{2}$ & $\left(40.22^{+1.38}_{-1.22}\right)^\circ$ & $\left(40.22^{+1.38}_{-1.22}\right)^\circ$ & $\left(40.22^{+1.38}_{-1.22}\right)^\circ$ & $\left(40.22^{+1.38}_{-1.22}\right)^\circ$ & $\left(40.22^{+1.38}_{-1.22}\right)^\circ$ & $A_{4}$\\
        \hline
        P6 & $\theta_{1}+\vartheta_{1}$ & $\left(36.30^{+1.23}_{-1.18}\right)^\circ$ & $\left(38.00^{+1.08}_{-1.13}\right)^\circ$ & $\left(41.74^{+1.05}_{-1.00}\right)^\circ$ & $\left(44.33^{+0.99}_{-0.92}\right)^\circ$ & $\left(45.05^{+0.98}_{-0.91}\right)^\circ$ & $B_{4}$\\
        & $\theta_{2}+\vartheta_{2}$ & $\left(32.48^{+1.52}_{-1.37}\right)^\circ$ & $\left(33.76^{+1.41}_{-1.26}\right)^\circ$ & $\left(36.68^{+1.27}_{-1.13}\right)^\circ$ & $\left(38.79^{+1.25}_{-1.12}\right)^\circ$ & $\left(39.38^{+1.26}_{-1.13}\right)^\circ$ & $B_{>5}$\\
        \hline
        P7 & $\theta_{1}+\vartheta_{1}$ & $\left(58.35^{+1.09}_{-0.95}\right)^\circ$ & $\left(59.59^{+1.06}_{-0.93}\right)^\circ$ & $\left(62.83^{+1.06}_{-0.93}\right)^\circ$ & $\left(65.60^{+1.20}_{-1.06}\right)^\circ$ & $\left(66.45^{+1.27}_{-1.12}\right)^\circ$ & $B_{>5}$\\
        & $\theta_{2}+\vartheta_{2}$ & $\left(70.39^{+1.76}_{-1.69}\right)^\circ$ & $\left(68.39^{+1.71}_{-1.63}\right)^\circ$ & $\left(64.08^{+1.56}_{-1.46}\right)^\circ$ & $\left(61.18^{+1.40}_{-1.31}\right)^\circ$ & $\left(60.39^{+1.35}_{-1.27}\right)^\circ$ & $B_{>5}$\\
        \hline
        P8 & $\theta_{1}+\vartheta_{1}$ & $\left(50.83^{+1.17}_{-1.05}\right)^\circ$ & $\left(51.44^{+1.20}_{-1.08}\right)^\circ$ & $\left(53.20^{+1.27}_{-1.15}\right)^\circ$ & $\left(54.90^{+1.41}_{-1.27}\right)^\circ$ & $\left(55.46^{+1.48}_{-1.32}\right)^\circ$ & $B_{>5}$\\
        & $\theta_{2}+\vartheta_{2}$ & $\left(29.05^{+1.23}_{-1.15}\right)^\circ$ & $\left(30.56^{+1.13}_{-1.05}\right)^\circ$ & $\left(34.27^{+1.02}_{-0.94}\right)^\circ$ & $\left(37.22^{+1.07}_{-0.99}\right)^\circ$ & $\left(38.09^{+1.10}_{-1.01}\right)^\circ$ & $B_{>5}$\\
        \hline
        P9 & $\theta_{1}+\vartheta_{1}$ & $\left(33.74^{+1.05}_{-1.04}\right)^\circ$ & $\left(35.55^{+1.02}_{-1.02}\right)^\circ$ & $\left(39.80^{+0.93}_{-0.93}\right)^\circ$ & $\left(43.04^{+0.89}_{-0.87}\right)^\circ$ & $\left(43.98^{+0.89}_{-0.86}\right)^\circ$ & $B_{>5}$\\
        & $\theta_{2}+\vartheta_{2}$ & $\left(43.43^{+1.35}_{-1.21}\right)^\circ$ & $\left(44.06^{+1.36}_{-1.21}\right)^\circ$ & $\left(45.85^{+1.27}_{-1.12}\right)^\circ$ & $\left(47.56^{+1.57}_{-1.40}\right)^\circ$ & $\left(48.12^{+1.64}_{-1.46}\right)^\circ$ & $B_{1}$\\
        \bottomrule
      \end{tabular}
      \end{ruledtabular}
\end{table*}
\endgroup

Here, we distinguish two types of the behavior of the sums by symbols
A and B, respectively:
\begin{description}
\item[Type A] The values are independent of the lepton {\it CP}-violating phase $\phi_3$.
\item[Type B] The values vary with the variation of the lepton {\it CP}-violating phase $\phi_3$.
\end{description}
The subscripts of each type represent the error limit. (In the case
of Type B, we classify the deviations only by the values in the condition
$\phi_3=90^\circ$.) For example, $A_{3}$ represents that the sum in Type A
deviates from $45^\circ$ with an error between $2\sigma$ and
$3\sigma$; $B_{>5}$ means that the sum with $\phi_3=90^\circ$ in
Type B deviates from $45^\circ$ with an error larger than $5\sigma$.

All the relations in Type A are relatively more consistent with the prediction of QLC,
while the phase-dependent property of Type B relations adds
complexities. Moreover, many Type B relations in
Table~\ref{complementarity} largely deviate from expectations. We
remind readers to pay special attention to the P1, P7, and P8
schemes. In P1, QLC1 are obviously in Type A and are close to the
expected value $45^\circ$, which in fact is a major cause leading to
the hypothesis of QLC. However, with the latest global fit
data~\cite{lepglobalfit}, QLC2 in P1 deviates from expectations with
an error larger than $3\sigma$ and thus even in the P1 scheme, QLC2
may not be good relations, which is obscured by relatively less
accurate data before. In P7 and P8, all QLC relations are far beyond
error limits, no matter what value of $\phi_3$ we choose, and thus
are hardly desired relations. Therefore, we see that the validation
of QLC in some schemes significantly depends on the lepton
{\it CP}-violating phase $\phi_3$ we choose, and in the P7 and the P8
schemes, QLC can never be satisfied. This dependence of QLC on the
choices of schemes and lepton {\it CP}-violating phase was sometimes
ignored by previous works.

Since the QLC relations are originally observed in the standard
CK scheme, this phase-dependent property and the generally
phase-dependent result of QLC in the other eight schemes remind us
to be cautious on the generalization of QLC from the CK scheme to
the other eight schemes. When considering such generalizations,
careful inspections and justifications should be carried out. In
addition, Jarlskog has pointed out a few of the uncertainties that
could invalidate the QLC analyses~\cite{jarlskog05}, which also
reminds us to carefully treat QLC relations. An alternative
way of avoiding such generalizations is to use some scheme-independent
forms of QLC relations. One example is to analyze QLC relations in
the form of matrix elements~\cite{hexiaogang,Zhang:2012xu,QLCnine}. Since we have figured
out the dependence of QLC relations on the {\it CP}-violating phase,
experimental results on the lepton {\it CP}-violating phase measured in the
future will be helpful in analyzing QLC in the other eight schemes.

\section{\label{sec4}SELF-COMPLEMENTARITY}
\begin{figure*}
        \centering
        \subfigure[~P1]{
          \includegraphics[width=5.8cm]{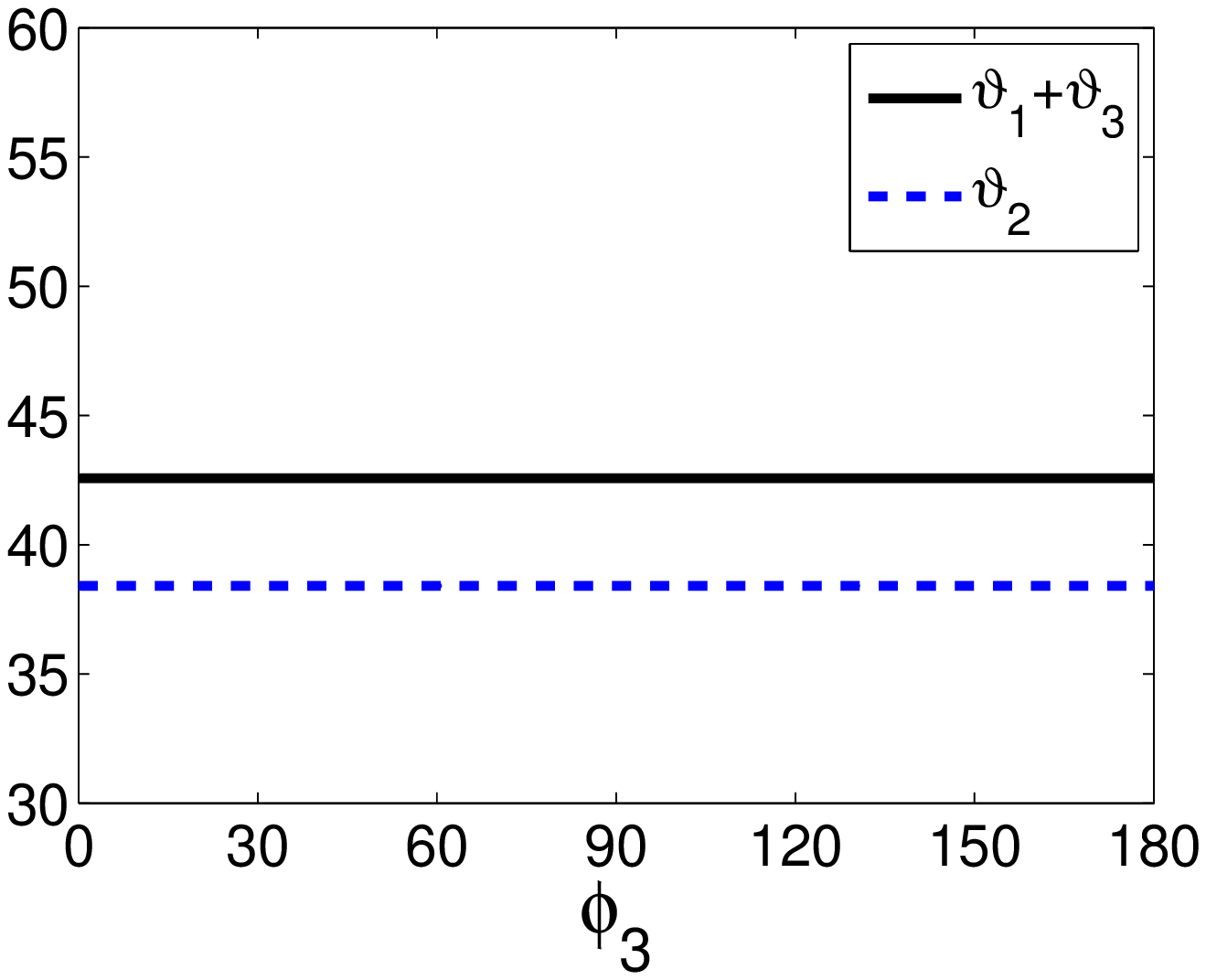}}
        \subfigure[~P2]{
          \includegraphics[width=5.8cm]{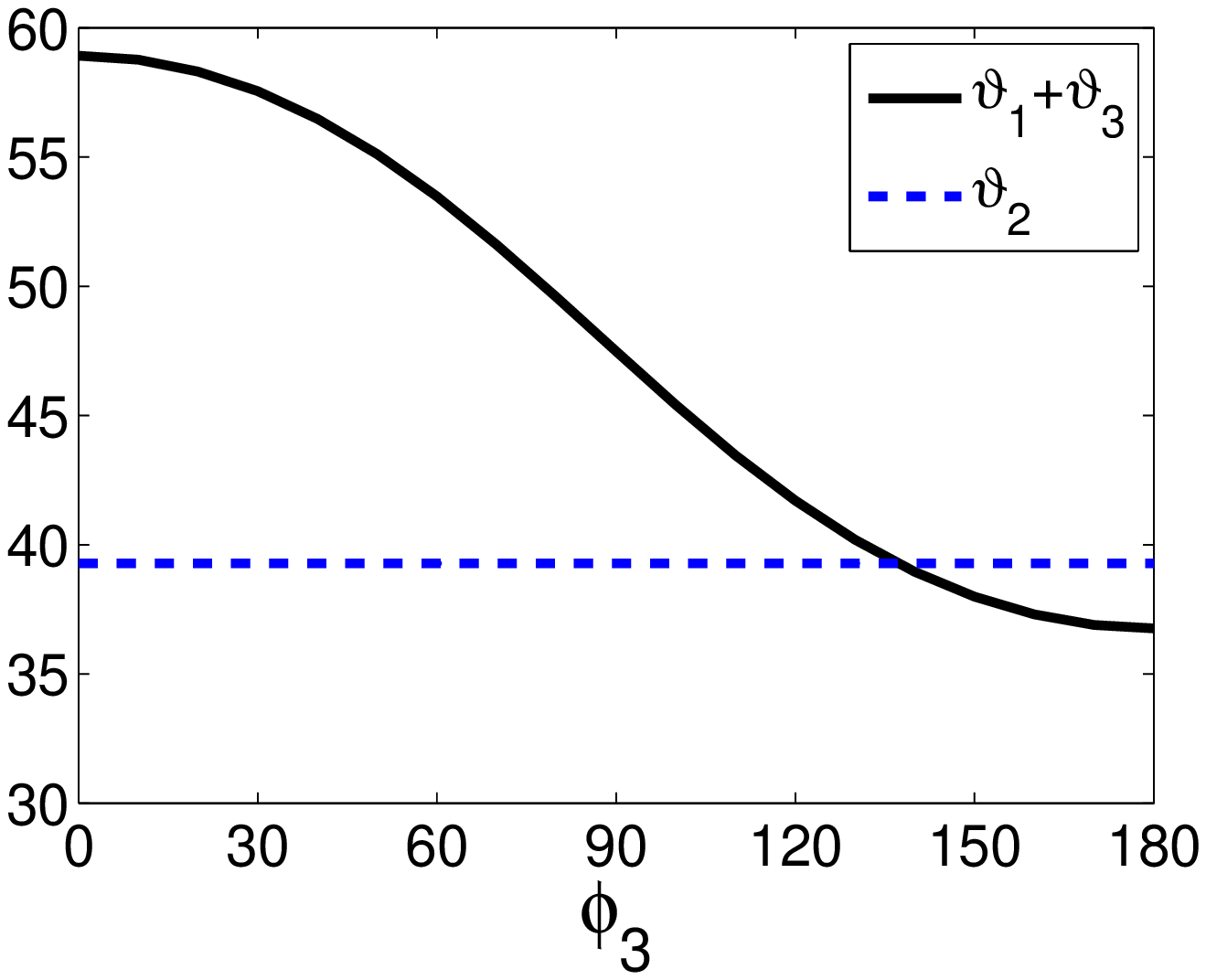}}
        \subfigure[~P3]{
          \includegraphics[width=5.8cm]{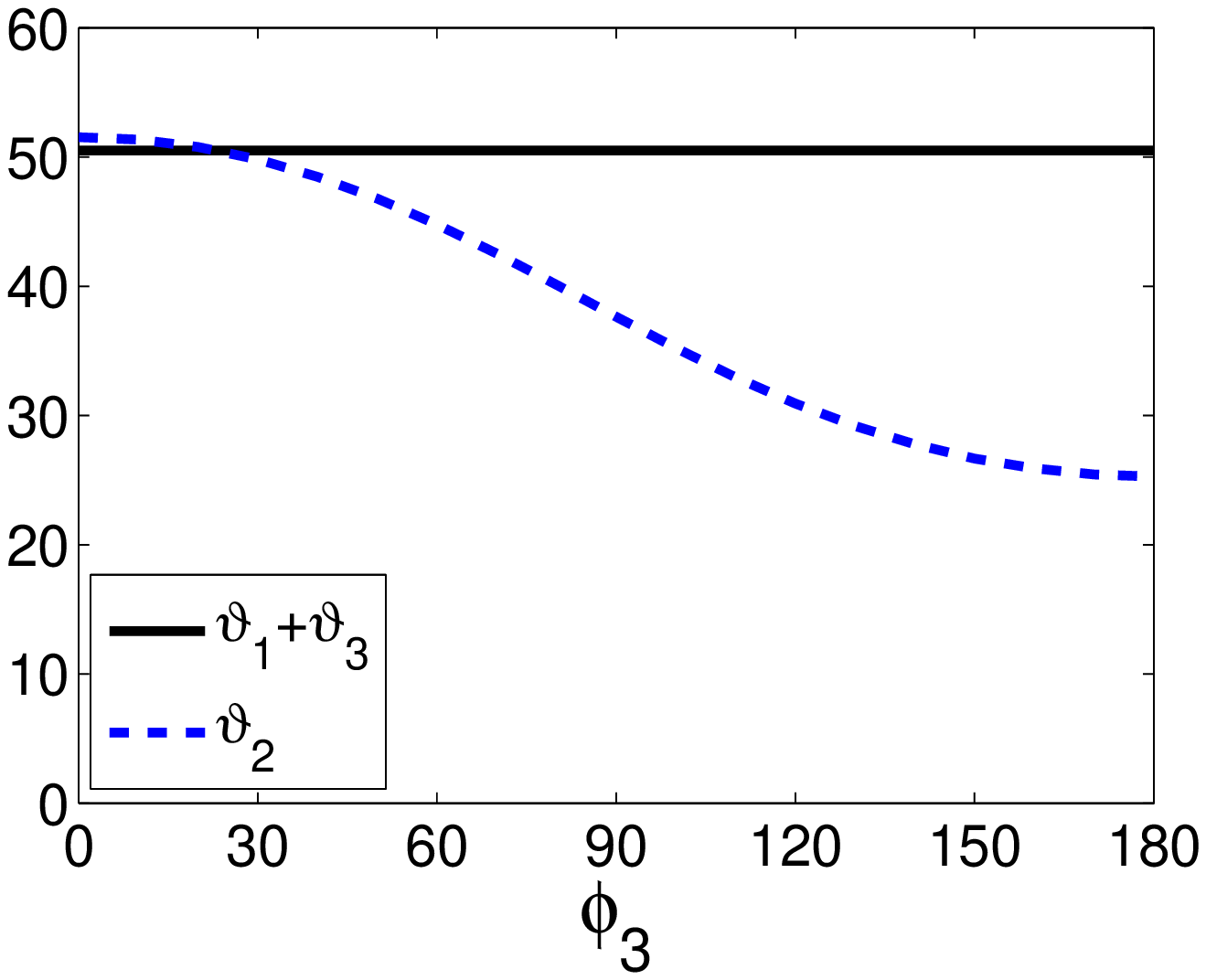}}
        \subfigure[~P4]{
          \includegraphics[width=5.8cm]{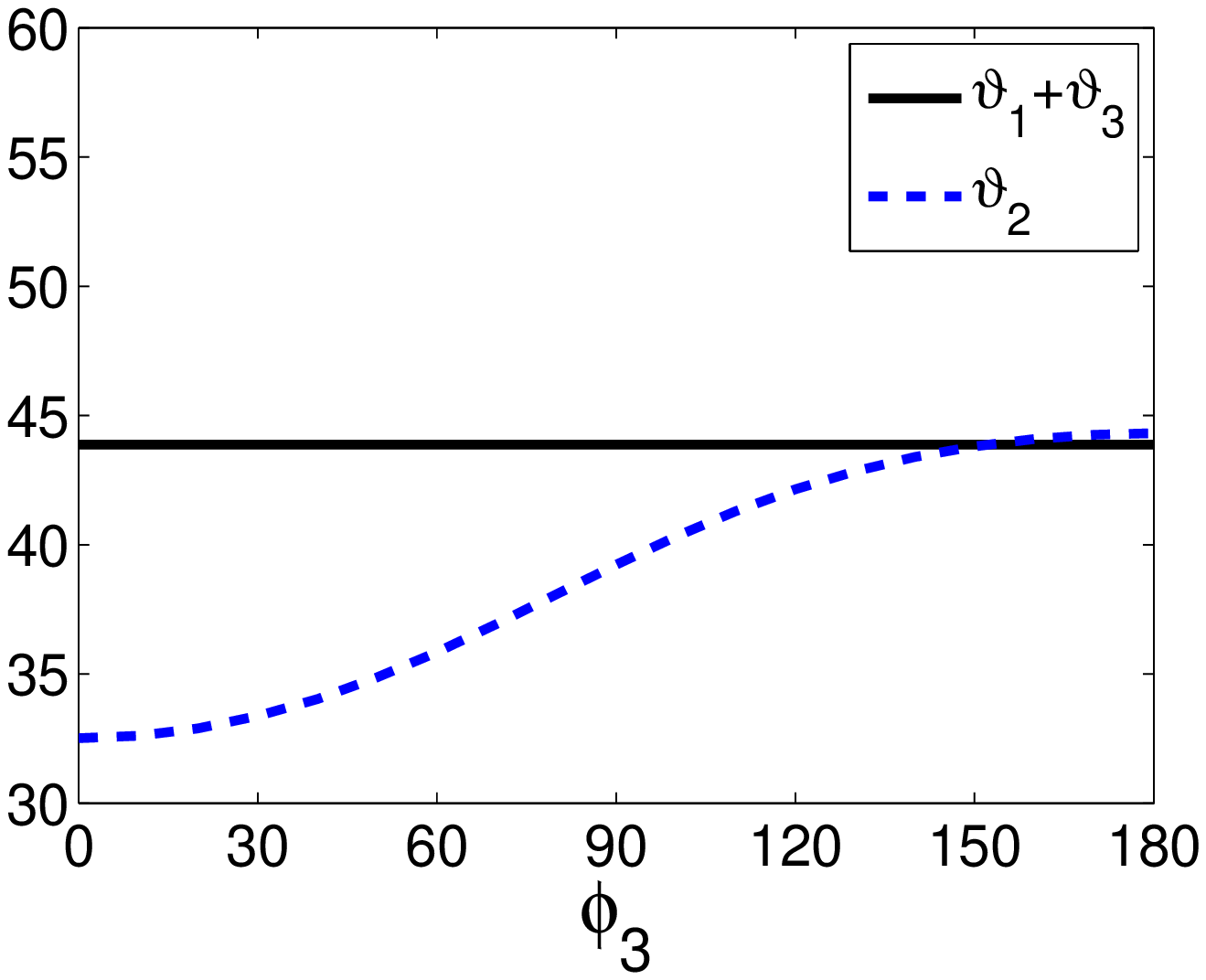}}
        \subfigure[~P5]{
          \includegraphics[width=5.8cm]{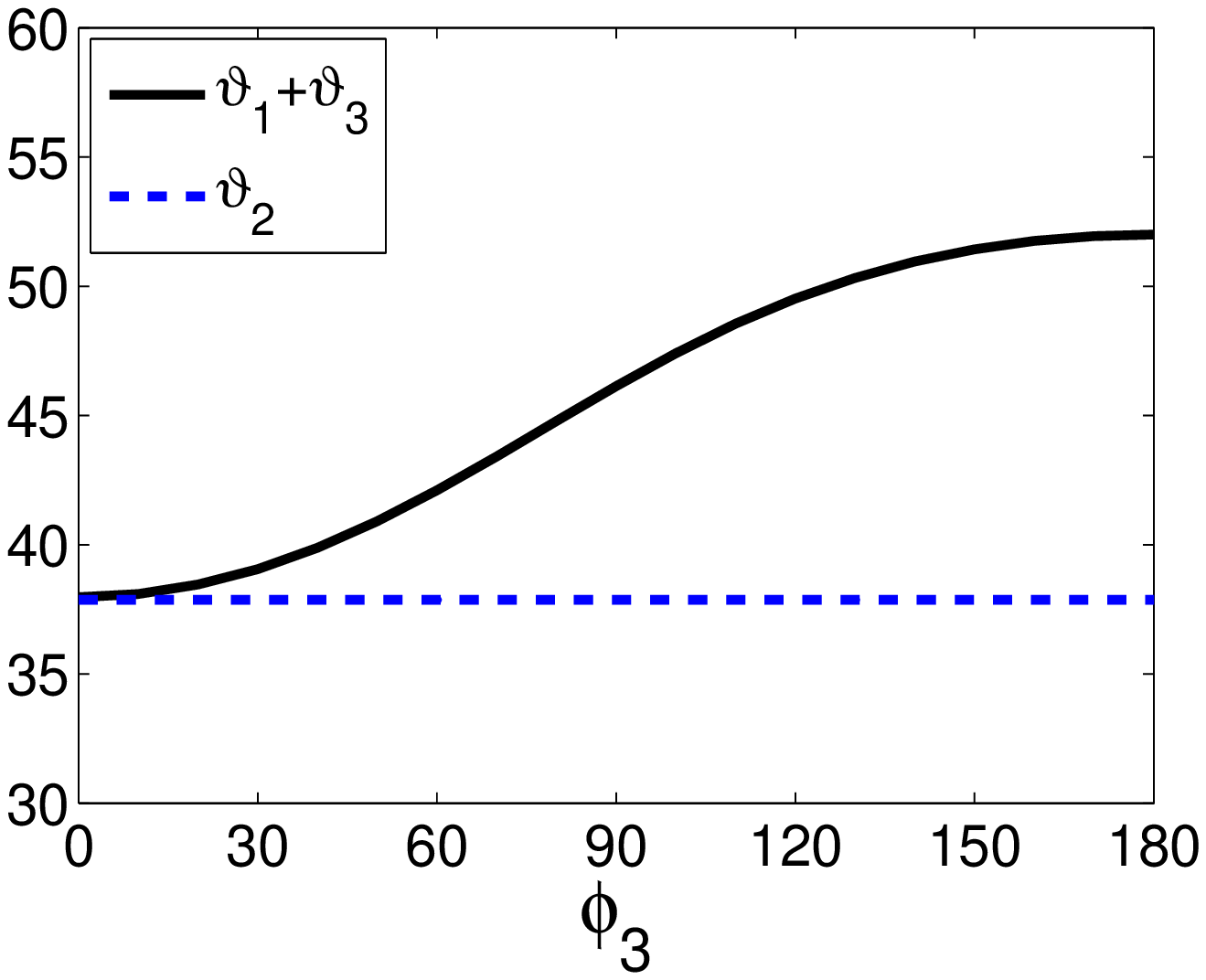}}
        \subfigure[~P6]{
          \includegraphics[width=5.8cm]{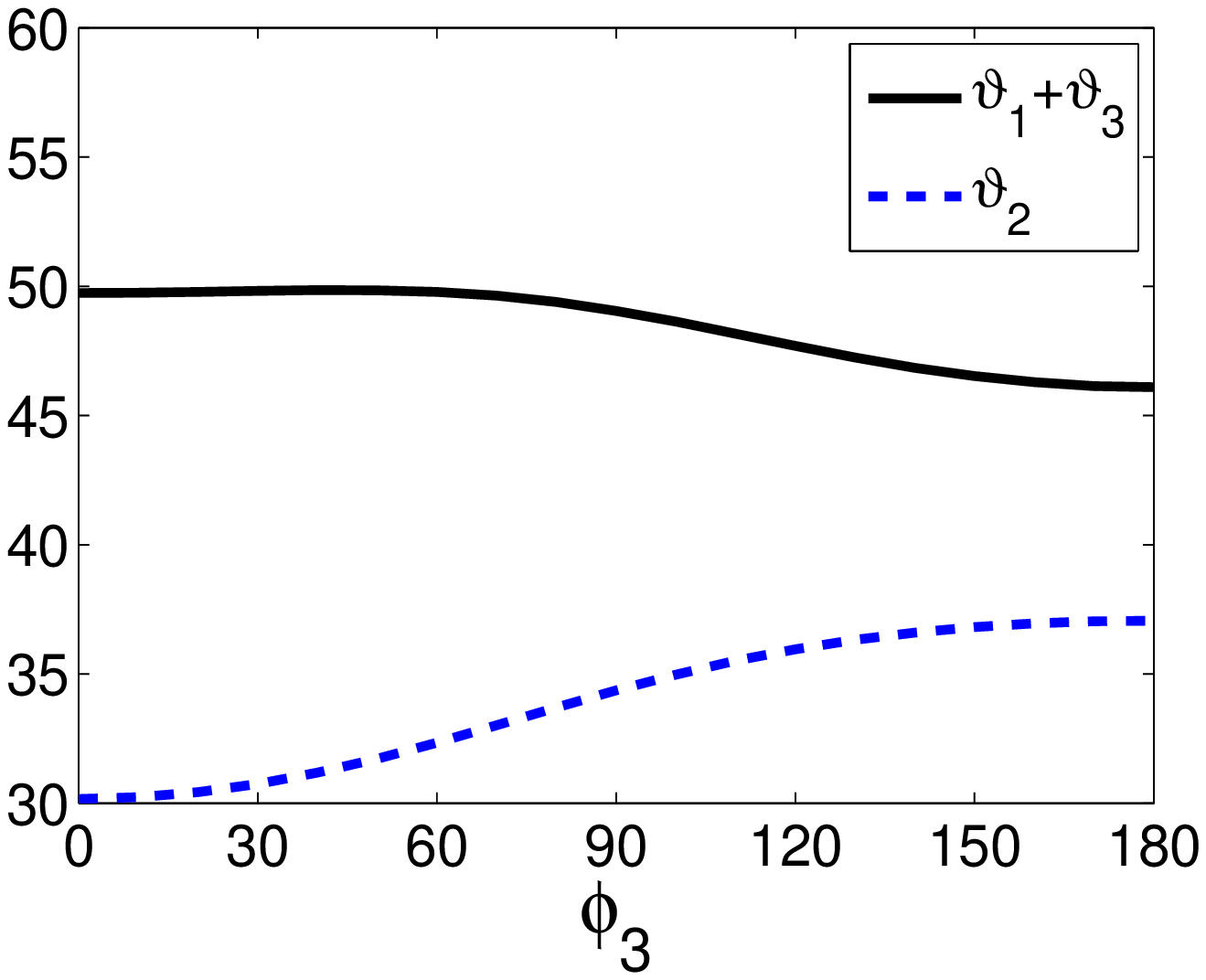}}
        \subfigure[~P7]{
          \includegraphics[width=5.8cm]{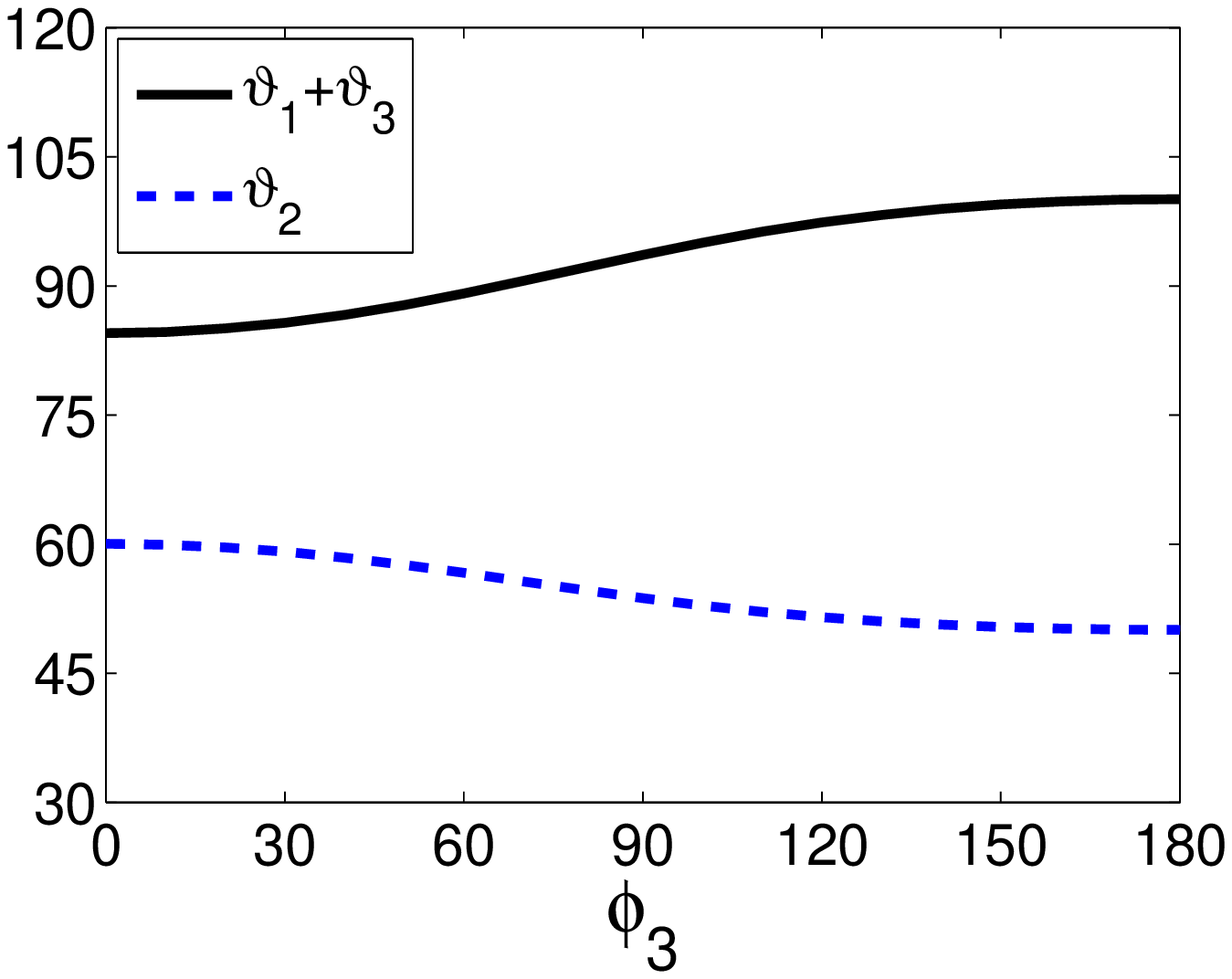}}
        \subfigure[~P8]{
          \includegraphics[width=5.8cm]{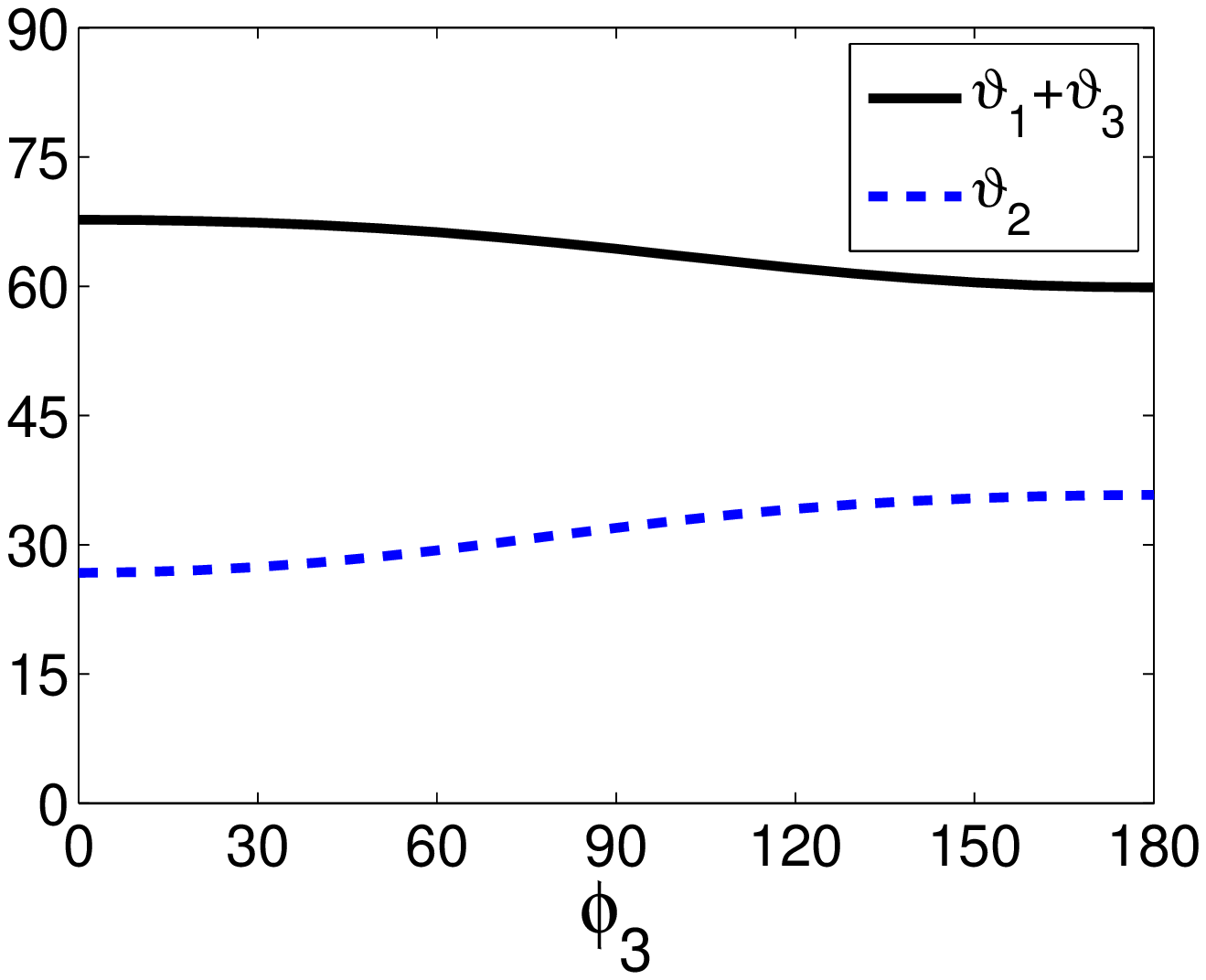}}
        \subfigure[~P9]{
          \includegraphics[width=5.8cm]{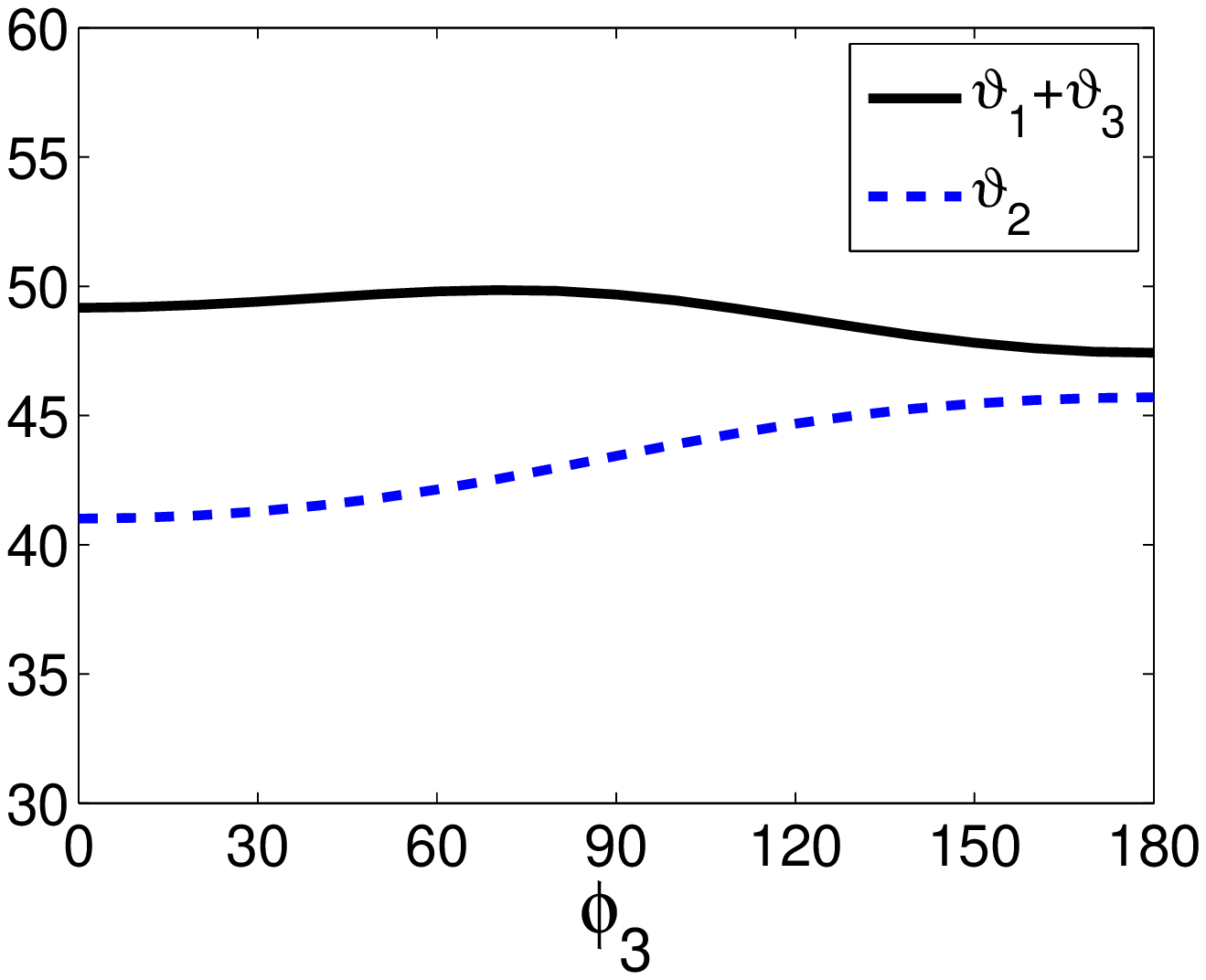}}
        \caption{The self-complementarity among lepton mixing angles [All the values are in the unit of degree~(${}^\circ$).]}
        \label{fig4}
    \end{figure*}

SC relations of lepton mixing angles are examined similarly as QLC, with the results shown in Fig.~\ref{fig4}. More detailed results with errors are provided in Table~\ref{scomplementarity}.
\begingroup
\squeezetable
\begin{table*}
      \caption{The self-complementarity among lepton mixing angles}\label{scomplementarity}
      \begin{ruledtabular}
      \begin{tabular}{cccccccc}
        \toprule
        & SC & $(\phi_3=0^\circ)$ & $(\phi_3=45^\circ)$ & $(\phi_3=90^\circ)$ & $(\phi_3=135^\circ)$ & $(\phi_3=180^\circ)$ & Type\\
        \hline
        P1 & $\vartheta_{12}+\vartheta_{13}$ & $\left(42.58^{+1.20}_{-1.11}\right)^\circ$ & $\left(42.58^{+1.20}_{-1.11}\right)^\circ$ & $\left(42.58^{+1.20}_{-1.11}\right)^\circ$ & $\left(42.58^{+1.20}_{-1.11}\right)^\circ$ & $\left(42.58^{+1.20}_{-1.11}\right)^\circ$ & $A_{3}$\\
        & $\vartheta_{23}$ & $\left(38.41^{+1.40}_{-1.24}\right)^\circ$ & $\left(38.41^{+1.40}_{-1.24}\right)^\circ$ & $\left(38.41^{+1.40}_{-1.24}\right)^\circ$ & $\left(38.41^{+1.40}_{-1.24}\right)^\circ$ & $\left(38.41^{+1.40}_{-1.24}\right)^\circ$ & \\
        \hline
        P2 & $\vartheta_{1}+\vartheta_{3}$ & $\left(58.92^{+1.56}_{-1.53}\right)^\circ$ & $\left(55.83^{+1.41}_{-1.38}\right)^\circ$ & $\left(47.48^{+1.36}_{-1.31}\right)^\circ$ & $\left(39.53^{+1.54}_{-1.47}\right)^\circ$ & $\left(36.76^{+1.58}_{-1.50}\right)^\circ$ & $B_{2}$\\
        & $\vartheta_{2}$ & $\left(39.28^{+1.37}_{-1.21}\right)^\circ$ & $\left(39.28^{+1.37}_{-1.21}\right)^\circ$ & $\left(39.28^{+1.37}_{-1.21}\right)^\circ$ & $\left(39.28^{+1.37}_{-1.21}\right)^\circ$ & $\left(39.28^{+1.37}_{-1.21}\right)^\circ$ & \\
        \hline
        P3 & $\vartheta_{1}+\vartheta_{3}$ & $\left(50.51^{+1.39}_{-1.34}\right)^\circ$ & $\left(50.51^{+1.39}_{-1.34}\right)^\circ$ & $\left(50.51^{+1.39}_{-1.34}\right)^\circ$ & $\left(50.51^{+1.39}_{-1.34}\right)^\circ$ & $\left(50.51^{+1.39}_{-1.34}\right)^\circ$ & $A_{5}$\\
        & $\vartheta_{2}$ & $\left(51.54^{+1.62}_{-1.51}\right)^\circ$ & $\left(47.67^{+1.60}_{-1.45}\right)^\circ$ & $\left(37.64^{+1.58}_{-1.40}\right)^\circ$ & $\left(28.44^{+1.64}_{-1.48}\right)^\circ$ & $\left(25.28^{+1.64}_{-1.48}\right)^\circ$ & \\
        \hline
        P4 & $\vartheta_{1}+\vartheta_{3}$ & $\left(43.88^{+1.23}_{-1.14}\right)^\circ$ & $\left(43.88^{+1.23}_{-1.14}\right)^\circ$ & $\left(43.88^{+1.23}_{-1.14}\right)^\circ$ & $\left(43.88^{+1.23}_{-1.14}\right)^\circ$ & $\left(43.88^{+1.23}_{-1.14}\right)^\circ$ & $A_{1}$\\
        & $\vartheta_{2}$ & $\left(32.51^{+1.46}_{-1.30}\right)^\circ$ & $\left(34.43^{+1.34}_{-1.19}\right)^\circ$ & $\left(39.23^{+1.23}_{-1.09}\right)^\circ$ & $\left(43.13^{+1.37}_{-1.22}\right)^\circ$ & $\left(44.31^{+1.46}_{-1.30}\right)^\circ$ & \\
        \hline
        P5 & $\vartheta_{1}+\vartheta_{3}$ & $\left(37.97^{+1.36}_{-1.28}\right)^\circ$ & $\left(40.37^{+1.30}_{-1.23}\right)^\circ$ & $\left(46.14^{+1.21}_{-1.14}\right)^\circ$ & $\left(50.66^{+1.29}_{-1.21}\right)^\circ$ & $\left(52.00^{+1.36}_{-1.28}\right)^\circ$ & $B_{1}$\\
        & $\vartheta_{2}$ & $\left(37.86^{+1.38}_{-1.22}\right)^\circ$ & $\left(37.86^{+1.38}_{-1.22}\right)^\circ$ & $\left(37.86^{+1.38}_{-1.22}\right)^\circ$ & $\left(37.86^{+1.38}_{-1.22}\right)^\circ$ & $\left(37.86^{+1.38}_{-1.22}\right)^\circ$ & \\
        \hline
        P6 & $\vartheta_{1}+\vartheta_{3}$ & $\left(49.74^{+1.50}_{-1.42}\right)^\circ$ & $\left(49.85^{+1.47}_{-1.39}\right)^\circ$ & $\left(49.05^{+1.42}_{-1.33}\right)^\circ$ & $\left(47.04^{+1.44}_{-1.32}\right)^\circ$ & $\left(46.09^{+1.43}_{-1.31}\right)^\circ$ & $B_{4}$\\
        & $\vartheta_{2}$ & $\left(30.16^{+1.52}_{-1.37}\right)^\circ$ & $\left(31.44^{+1.41}_{-1.26}\right)^\circ$ & $\left(34.36^{+1.27}_{-1.13}\right)^\circ$ & $\left(36.47^{+1.25}_{-1.12}\right)^\circ$ & $\left(37.06^{+1.26}_{-1.13}\right)^\circ$ & \\
        \hline
        P7 & $\vartheta_{1}+\vartheta_{3}$ & $\left(84.52^{+1.96}_{-1.92}\right)^\circ$ & $\left(87.19^{+1.84}_{-1.80}\right)^\circ$ & $\left(93.62^{+1.68}_{-1.64}\right)^\circ$ & $\left(98.63^{+1.72}_{-1.66}\right)^\circ$ & $\left(100.10^{+1.76}_{-1.68}\right)^\circ$ & $B_{>5}$\\
        & $\vartheta_{2}$ & $\left(60.03^{+1.76}_{-1.69}\right)^\circ$ & $\left(58.03^{+1.71}_{-1.63}\right)^\circ$ & $\left(53.72^{+1.56}_{-1.46}\right)^\circ$ & $\left(50.82^{+1.40}_{-1.31}\right)^\circ$ & $\left(50.03^{+1.35}_{-1.27}\right)^\circ$ & \\
        \hline
        P8 & $\vartheta_{1}+\vartheta_{3}$ & $\left(67.72^{+1.63}_{-1.46}\right)^\circ$ & $\left(66.93^{+1.68}_{-1.51}\right)^\circ$ & $\left(64.35^{+1.85}_{-1.64}\right)^\circ$ & $\left(61.16^{+2.02}_{-1.80}\right)^\circ$ & $\left(59.85^{+2.07}_{-1.84}\right)^\circ$ & $B_{>5}$\\
        & $\vartheta_{2}$ & $\left(26.73^{+1.23}_{-1.15}\right)^\circ$ & $\left(28.24^{+1.13}_{-1.05}\right)^\circ$ & $\left(31.95^{+1.02}_{-0.94}\right)^\circ$ & $\left(34.90^{+1.07}_{-0.99}\right)^\circ$ & $\left(35.77^{+1.10}_{-1.01}\right)^\circ$ & \\
        \hline
        P9 & $\vartheta_{1}+\vartheta_{3}$ & $\left(49.17^{+1.42}_{-1.36}\right)^\circ$ & $\left(49.62^{+1.42}_{-1.36}\right)^\circ$ & $\left(49.68^{+1.45}_{-1.37}\right)^\circ$ & $\left(48.25^{+1.53}_{-1.42}\right)^\circ$ & $\left(47.42^{+1.55}_{-1.43}\right)^\circ$ & $B_{4}$\\
        & $\vartheta_{2}$ & $\left(41.01^{+1.35}_{-1.21}\right)^\circ$ & $\left(41.64^{+1.36}_{-1.21}\right)^\circ$ & $\left(43.43^{+1.27}_{-1.12}\right)^\circ$ & $\left(45.14^{+1.57}_{-1.40}\right)^\circ$ & $\left(45.70^{+1.64}_{-1.46}\right)^\circ$ & \\
        \bottomrule
      \end{tabular}\end{ruledtabular}
\end{table*}
\endgroup
The classification into two types and the definition of subscripts follow our treatment with QLC, and the values of $\vartheta_2$ are included for reference. Here, we also remind readers to pay attention to P1, P7, and P8 results. The former one fits relatively well, while the latter two are hardly desired relations. Similarly, we should be cautious about the generalization of SC relations to the other eight schemes, and it is clear that experimental results of the lepton {\it CP}-violating phase will be helpful in the examination of SC in the other eight schemes.
\section{\label{sec5}CP-VIOLATING PHASES}
\subsection{Analysis of results}
The variation of lepton {\it CP}-violating phases in different schemes
along with the variation of the {\it CP}-violating phase $\phi_3$ is not
trivial. From Fig.~\ref{fig1}, the relationship between $\phi_1\sim\phi_5$ and $\phi_3$ is quite
close to linear dependence.
More interesting are the nonmonotonous
relations between $\phi_6\sim\phi_9$ and $\phi_3$. From
Fig.~\ref{fig1} we guess that the correlation functions in
P6$\sim$P9 possess extremums with respect to $\phi_3$, with the extremums reached when
$\phi_3\simeq90^\circ$. Actually, through calculations we know that
the extremums in P6$\sim$P9 are reached when $\phi_3$ approaches
approximately $87.0^\circ$, $78.4^\circ$, $92.0^\circ$, and
$89.7^\circ$, respectively. Here the last one in P9 deserves attention, because it is quite
close to $90^\circ$.

Furthermore, Fig.~\ref{fig1} implies a way of redefinition of
{\it CP}-violating phases. By substituting $(180^\circ-\phi)$ for the
present {\it CP}-violating phases $\phi$ in P1, P8, and P9 schemes, we can
unify P1$\sim$P5 with the common property of similar
quasilinear correlation functions between their {\it CP}-violating phases
and $\phi_3$. The others, P6$\sim$P9 are
also united in this way, holding the same property of the existence
of a maximum value in similar correlation functions. This result
also indicates that without any knowledge on lepton {\it CP}-violating
phases, possible values of {\it CP}-violating phases in P6$\sim$P9 schemes
are already restricted by our known values of mixing angles.
Meanwhile, possible values of {\it CP}-violating phases in P1$\sim$P5
schemes are not restricted with current experimental data.
\subsection{Maximal CP violation}
It is necessary to clarify the meaning of ``maximal {\it CP} violation''
here. ``Maximal {\it CP} violation'' is defined as the case when the
magnitude of the scheme-independent Jarlskog invariant in the quark
sector takes its maximal value. However, there are ambiguities on
the choice of variables when we consider the meaning of ``maximal.''
Originally, all the four parameters are viewed as variables, but such
a maximized Jarlskog quantity is excluded in the quark sector
experimentally. Now, it is prevalent to view only the {\it CP}-violating
phase in each scheme as the variable with the mixing angles fixed.
For instance, the analysis of maximal {\it CP} violation is usually
carried out in the P3 scheme, where we regard mixing angles in P3 as
constant and choose the {\it CP}-violating phase $\phi_3$ as a variable.
In this interpretation of ``maximal {\it CP} violation,'' together with
our previous analysis that {\it CP}-violating phases in P1$\sim$P5 schemes
are unrestricted, the meaning of maximal {\it CP} violation is, in fact,
setting {\it CP}-violating phase in any one of P1$\sim$P5 schemes to be
$90^\circ$.

There is a conjecture that maximal {\it CP} violation is simultaneously
satisfied in both the quark and the lepton
sectors~\cite{Zhang:2012ys}. Taking the quark sector into
consideration, from Table~\ref{tab1} we easily recognize that
the P2 and P3 schemes in the quark sector possess large
{\it CP}-violating phases that equal $90^\circ$ within an error of
$1\sigma$, while in other schemes the {\it CP}-violating phases are far
from $90^\circ$. Therefore, the P2 and P3 schemes are the
favored ones when considering simultaneous maximal {\it CP} violation in
both the quark and the lepton sector.

\subsection{Empirical relations}
Finally, some empirical relations of the quark {\it CP}-violating phases in different schemes are explored. To better illustrate the results, we use the {\it CP}-violating phase redefinition suggested above, i.e., substituting $(180^\circ-\phi)$ for $\phi$ in P1, P8 and P9. For convenience, these nine {\it CP}-violating phases in the quark sector are relisted in Table~\ref{CP quark} with our redefinitions.
\begin{table*}
      \caption{The quark {\it CP}-violating phases in nine schemes}\label{CP quark}
\begin{scriptsize}
      \begin{ruledtabular}
      \begin{tabular}{ccccccccc}
        \toprule
        $\phi_1$ & $\phi_2$ & $\phi_3$ & $\phi_4$ & $\phi_5$ & $\phi_6$ & $\phi_7$ & $\phi_8$ & $\phi_9$\\
        \hline
        $\left(110.90^{+3.85}_{-2.02}\right)^\circ$ &
        $\left(89.69^{+2.29}_{-3.95}\right)^\circ$ &
        $\left(89.29^{+3.99}_{-2.33}\right)^\circ$ &
        $\left(111.95^{+3.82}_{-2.02}\right)^\circ$ &
        $\left(111.94^{+3.85}_{-2.02}\right)^\circ$ &
        $\left(22.72^{+1.25}_{-1.18}\right)^\circ$ &
        $\left(1.08^{+0.06}_{-0.06}\right)^\circ$ &
        $\left(22.69^{+1.25}_{-1.18}\right)^\circ$ &
        $\left(21.68^{+1.20}_{-1.13}\right)^\circ$\\
        \bottomrule
      \end{tabular}
      \end{ruledtabular}
\end{scriptsize}
\end{table*}
Some empirical relations we can easily read out are listed here:
\begin{eqnarray}
        &&\phi_1\sim\phi_4\sim\phi_5,\label{1}\\
        &&\phi_2\sim\phi_3\sim90^\circ,\label{2}\\
        &&\phi_6\sim\phi_8\sim\phi_9,\label{3}\\
        &&\phi_7\sim0^\circ.\label{4}
\end{eqnarray}
In fact, (\ref{1}) are satisfied by the similarities between their mixing angles. From Table~\ref{tab1}, we have these relations approximately~(here $\theta_{i(j)}$ represents $\theta_i$ in P$j$ scheme):
\begin{eqnarray}
        &&\theta_{12(1)}=\theta_{1(4)}=\theta_{1(5)},\\
        &&\theta_{23(1)}=\theta_{2(4)}=\theta_{2(5)},\\
        &&\theta_{13(1)}=\theta_{3(4)}=\theta_{3(5)}.
\end{eqnarray}
Then using the scheme-independent Jarlskog invariant, we get
\begin{eqnarray}
        &&1=\frac{J_1}{J_4}=\frac{s_{12(1)}s_{23(1)}s_{13(1)}c_{12(1)}c_{23(1)}c^2_{13(1)}\sin\phi_1}
        {s_{1(4)}s_{2(4)}s_{3(4)}c^2_{1(4)}c_{2(4)}c_{3(4)}\sin\phi_4}\nonumber\\
        &&\simeq\frac{c_{13(1)}\sin\phi_1}{c_{1(4)}\sin\phi_4}\simeq\frac{\sin\phi_1}{\sin\phi_4},\\
        &&1=\frac{J_4}{J_5}=\frac{s_{1(4)}s_{2(4)}s_{3(4)}c^2_{1(4)}c_{2(4)}c_{3(4)}\sin\phi_4}
        {s_{1(5)}s_{2(5)}s_{3(5)}c_{1(5)}c^2_{2(5)}c_{3(5)}\sin\phi_5}\nonumber\\
        &&\simeq\frac{c_{1(4)}\sin\phi_4}{c_{2(5)}\sin\phi_5}\simeq\frac{\sin\phi_4}{\sin\phi_5},
\end{eqnarray}
justifying the relation (\ref{1}). By the same way, (\ref{3}) is justified through the similarities among mixing angles in P6, P8, and P9.

Relation (\ref{2}) states possible maximal {\it CP} violation as we
discussed before, and relation (\ref{4}) merely reflects the
relative largeness of the three mixing angles in P7. With the
existence of {\it CP} violation confirmed, $\phi_7$ cannot be
exactly $0^\circ$ though close to it~\cite{Gerard}.

We are willing to find out some similar empirical relations on lepton {\it CP}-violating phases. Unfortunately, similar relations cannot be easily found, because the lepton mixing angles are quite different from each other in different schemes, thus invalidating our method used for the quark sector.
\section{CONCLUSIONS}
From the results of Sec.~\ref{sec3} and Sec.~\ref{sec4}, the validation of QLC and SC depends on the choices of schemes and lepton {\it CP}-violating phases, and careful inspections should be carried out when we consider the generalization of QLC and SC from the standard CK scheme to the other eight schemes. On the issues of {\it CP}-violating phases, restrictions on lepton {\it CP}-violating phases in P6$\sim$P9 are recognized. Simultaneous maximal {\it CP} violation in both the quark and the lepton sector is possible in the P2 and P3 scheme. A redefinition of {\it CP}-violating phases for unification is suggested and some empirical relations on the quark {\it CP}-violating phases are explored. All of these results may enrich our knowledge of QLC, SC relations, and {\it CP}-violating phases, helping us understand the mystery of lepton mixing.

\begin{acknowledgments}
 This work is supported by the Principal Fund for Undergraduate Research at Peking
University. It is also partially supported by the National Natural
Science Foundation of China (Grants No.~11021092, No.~10975003,
No.~11035003, and No.~11120101004), by the National Fund for Fostering
Talents of Basic Science (Grants No.~J1030310), and by the
Research Fund for the Doctoral Program of Higher Education (China).
\end{acknowledgments}


\begin{thebibliography}{50}
\bibitem{PMNS}
    B.~Pontecorvo, Sov. Phys. JETP {\bf 26},  984 (1968);\\
    Z.~Maki, M.~Nakagawa and S.~Sakata, Prog. Theor. Phys.  {\bf 28},
    870 (1962).
\bibitem{CKM}
    N.~Cabibbo, Phys. Rev. Lett.  {\bf 10},  531 (1963);\\
    M.~Kobayashi and T.~Maskawa, Prog. Theor. Phys.  {\bf 49}, 652 (1973).
\bibitem{CK}
    L.L.~Chau and W.Y.~Keung,
    Phys. Rev. Lett. {\bf 53}, 1802 (1984).
\bibitem{KM}
    M.~Kobayashi and T.~Maskawa, Prog. Theor. Phys.  {\bf 49} 652 (1973).
\bibitem{koide}
    Y. Koide, Phys. Lett. B {\bf607}, 123 (2005).
\bibitem{Koide:2008yu}
    Y.~Koide and H.~Nishiura,
    Phys. Rev. D {\bf 79}, 093005 (2009).
\bibitem{boomerang}
    P.H.~Frampton and X.-G.~He, Phys. Lett. B {\bf688}, 67 (2010);
    P.H.~Frampton and X.-G.~He, Phys. Rev. D {\bf82}, 017301 (2010).
\bibitem{Li:2010ae}
    S.W.~Li and B.-Q.~Ma,
    Phys. Lett.  B {\bf 691}, 37 (2010).
\bibitem{qinnan}
    N.~Qin and B.-Q.~Ma, Phys. Lett. B 695, 194 (2011); N.~Qin and
    B.-Q.~Ma, Phys. Rev. D 83, 033006 (2011).
\bibitem{Ahn:2011it}
    Y.H.~Ahn, H.Y.~Cheng and S.~Oh,
    Phys. Lett.  B {\bf 701},  614 (2011).
\bibitem{nine}
    H. Fritzsch and Z.-Z. Xing,
    Phys. Rev.  D {\bf 57} 594 (1998).
\bibitem{Zheng10}
    Y.-j.~Zheng, Phys. Rev. D {\bf 81}, 073009 (2010).
\bibitem{Zhang:2012xu}
    X.~Zhang and B.-Q.~Ma,
    Phys. Lett. B {\bf 710}, 630 (2012).
\bibitem{QLCnine}
    X.~Zhang, Y.-j.~Zheng and B.-Q.~Ma,
    Phys. Rev.  D {\bf 85} 097301 (2012).
\bibitem{smirnov}
    A.~Y.~Smirnov, arXiv:hep-ph/0402264.
\bibitem{qlc}
    H.~Minakata and A.Y.~Smirnov, Phys. Rev. {\bf 70}, 073009 (2004).
\bibitem{raidal}
    M.~Raidal, Phys. Rev. Lett.  {\bf 93},  161801 (2004).
\bibitem{phenomenology}
    See, e.g.,
    P.~H.~Frampton and R.~N.~Mohapatra,
    J. High Energy Phys. {\bf 0501},  025 (2005);
    N.~Li and B.-Q.~Ma,
    Phys. Rev.  D {\bf 71}, 097301 (2005);
    S.~Antusch, S.~F.~King and R.~N.~Mohapatra,
    Phys. Lett.  B {\bf 618},  150 (2005);
    H.~Minakata,;
    J.~Ferrandis and S.~Pakvasa, Phys. Rev. D {\bf71}, 033004 (2005);
    S.K.~Kang, C.S.~Kim and J.~Lee, Phys. Lett. B {\bf619}, 129 (2005);
    M.A.~Schmidt and A.Y.~Smirnov, Phys. Rev. D {\bf74}, 113003 (2006);
    K.~A.~Hochmuth and W.~Rodejohann,
    Phys. Rev.  D {\bf 75}, 073001 (2007);
    F.~Plentinger, G.~Seidl and W.~Winter, Phys. Rev. D {\bf76}, 113003
    (2007);
    G.~Altarelli, F.~Feruglio and L.~Merlo,
    J. High Energy Phys. {\bf 0905}, 020 (2009).
\bibitem{Zheng:2011uz}
    Y.-j.~Zheng and B.-Q.~Ma,
    Eur. Phys. J. Plus {\bf 127}, 7 (2012).
\bibitem{Wolfenstein:1983yz}
    L.~Wolfenstein,
    Phys. Rev. Lett.  {\bf 51}, 1945 (1983).
\bibitem{pdg2012}
    J.~Beringer {\it et al.} (Particle Data Group),
    Phys. Rev.  D {\bf 86}, 010001 (2012).
\bibitem{jarlskog}
    C.~Jarlskog, Phys. Rev. Lett. {\bf 55} 1039 (1985);\\
    D.-d.~Wu,
    Phys. Rev.  D {\bf 33} 860 (1986);\\
    O.W.~Greenberg, Phys. Rev. D {\bf32} 1841 (1985).
\bibitem{lepglobalfit}
    G.~L.~Fogli, E.~Lisi, A.~Marrone, D.~Montanino, A.~Palazzo, and A.~Rotunno, Phys. Rev. D {\bf 86}, 013012 (2012).
\bibitem{Zhang:2012ys}
    X.~Zhang and B.-Q.~Ma,
    Phys. Lett. B {\bf 713}, 202 (2012);
    arXiv:1204.6604 [hep-ph].
\bibitem{T2K}
    K.~Abe {\it et al.}, Phys. Rev. Lett. {\bf 107} 041801 (2011),.
\bibitem{MINOS}
    P.Adamson {\it et al.}, Phys. Rev.  Lett. {\bf 107}, 181802 (2011).
\bibitem{jarlskog05}
    C.~Jarlskog
    Phys. Lett. B {\bf 625}, 63 (2005).
\bibitem{hexiaogang}
    G.-N.~Li, H.-H.~Lin and X.-G.~He,
    Phys. Lett. B {\bf 711}, 57 (2012).
\bibitem{Gerard}
See, also, Jean-Marc Gerard, arXiv:0811.0540.
\end{thebibliography}
\end{document}